\documentclass[12pt]{article}

\usepackage{amsthm,amsmath,amsfonts,amssymb}
\usepackage{graphicx}
\usepackage[authoryear]{natbib}
\usepackage[justification=raggedright]{subcaption}
\usepackage{xcolor}
\usepackage{afterpage}
\usepackage{placeins}
\usepackage{blindtext}
\usepackage{fancyhdr} 
\usepackage{multirow}
\usepackage{booktabs} 

\usepackage{setspace}
\usepackage[margin=1in]{geometry}  

\newcommand{\keywords}[1]{\textbf{Keywords:} #1}

\usepackage[colorlinks,citecolor=blue,urlcolor=blue,backref=page]{hyperref}

\newcommand{\Mymodel}{H-CLS}

\newtheorem{definition}{Definition}[section]

\title{Hyperbolic Latent Space Models for Network Embedding: Model Specification and Bayesian Inference}

\author{
Yiwei Gong$^{1}$, Anna L. Smith$^{2}$, Dena Asta$^{3}$, and Catherine A. Calder$^{1}$\\[0.5em]
{\small $^{1}$Department of Statistics and Data Sciences, The University of Texas at Austin}\\
{\small $^{2}$Department of Statistics, University of Kentucky}\\
{\small $^{3}$Department of Statistics, The Ohio State University}
}

\date{}

\begin{document}
\maketitle

\renewcommand{\thefootnote}{}
\footnotetext{Code available at \url{https://github.com/ywgej9/network_geometry}}
\renewcommand{\thefootnote}{\arabic{footnote}}

\begin{abstract}
Many real-world networks exhibit hierarchical, tree-like structure and heavy-tailed degree distributions, phenomena not readily captured by standard statistical models for network data. Extensions of the popular continuous latent space modeling framework have been proposed to accommodate such networks. Drawing on insights from statistical physics, continuous latent space models with underlying hyperbolic geometry have been proposed as a natural framework, probabilistically embedding nodes in a latent Riemannian manifold with constant negative curvature. Most statistical implementations, however, simplify the original physics-based model by omitting the ``temperature parameter," which controls the sharpness of the latent distance-to-probability mapping. We argue this omission is critical. We demonstrate that temperature is the fundamental parameter governing a network's tree-like topology, and that failing to infer it weakens model expressiveness. We formalize a Bayesian hyperbolic continuous latent space model with an unknown, learnable temperature parameter. We then develop two inferential procedures: a Hamiltonian Monte Carlo approach for rigorous posterior characterization and a scalable auto-encoding variational Bayes algorithm for large-scale networks. Through simulation and real data examples, we show that our model outperforms models with fixed temperature and misspecified Euclidean geometries in graph reconstruction tasks in most settings, confirming temperature is a crucial and inferable feature of complex networks. 
\end{abstract}


\section{Introduction}

Networks provide a natural mathematical framework for analyzing relational data across numerous scientific domains from sociology and biology to computer science and physics~\citep{guimera2004modeling, nr, bardoscia2021physics, camacho2018next}. These structures, composed of a set of entities (nodes) and the relationships or interactions (edges) between them, offer a powerful foundation for understanding complex systems. However, the statistical analysis of network data presents unique computational and methodological challenges that distinguish it from classical data structures. The inherent dependencies among observations violate the standard independent and identically distributed data assumption; for instance, the existence of two edges with a shared vertex may be correlated with the existence of a third edge connecting their other two endpoints, a phenomenon known as transitivity. Furthermore, with $N$ nodes, the number of potential relationships scales quadratically $O(N^2)$, leading to severe high-dimensionality issues even for networks of moderate size. These complexities necessitate a specialized class of statistical models capable of capturing the intricate dependency structure and providing a representation of the underlying generative mechanism.

To address these challenges, several families of statistical network models have been developed. Exponential random graph models (ERGMs), for instance, directly parameterize a graph's probability distribution as a function of network statistics, offering a powerful tool for testing specific hypotheses~\citep{wasserman1996logit, robins2007introduction, holland1981exponential}. Alternatively, stochastic block models (SBMs) and their variants assume a discrete latent structure, partitioning nodes into distinct communities or blocks where edge probabilities are uniform within and between blocks~\citep{abbe2018community, holland1983stochastic, lei2015consistency}. While highly effective for uncovering large-scale community structure, this discrete assumption can oversimplify the continuous social heterogeneity present in real-world systems. Mixed membership extensions~\citep{airoldi2008mixed} partially address this by allowing nodes to belong to multiple communities, but still rely on a finite set of discrete blocks. Latent space models~\citep{hoff2002latent, mccormick2015latent, smith2019geometry} provide a fully continuous alternative by embedding nodes into a low-dimensional latent space, typically Euclidean, where the probability of a connection is modeled as a function of the nodes' proximity. This approach provides an intuitive framework for visualizing network structure and naturally captures phenomena like homophily and transitivity.

The original latent space model~\citep{hoff2002latent} embeds nodes in a low-dimensional Euclidean space, providing an intuitive and interpretable framework for network analysis. However, this foundational approach faces limitations when modeling complex hierarchical structures common in real networks. To accommodate such organization, Euclidean models often require either the latent space to be of high dimension or increased model complexity. For instance, to capture heterogeneity and hierarchical structure within the Euclidean framework,~\cite{handcock2007model} proposed a model-based clustering for latent positions, using Gaussian mixture models in the latent space. This approach achieves flexibility through multivariate normal mixture components, creating local neighborhoods with different connectivity patterns. While effective, this requires estimating multiple cluster centers, covariance matrices, and mixture weights, substantially increasing model complexity. Similarly, work by~\cite{schweinberger2015local} on local dependence in ERGMs demonstrates that even in the ERGM framework, representing hierarchical dependencies requires high-dimensional parameters. These examples show that Euclidean continuous latent space models demand considerable model complexity to represent hierarchical networks. This motivates an alternative approach: rather than adding complexity to the model, we can change the underlying geometry. Non-Euclidean spaces, particularly hyperbolic manifolds, naturally encode the hierarchical structures that require extensive parameterization in Euclidean settings, achieving with simple two-dimensional representations what would otherwise demand complex model specifications.

This shift towards alternative geometries fundamentally reshapes how models represent network structure.~\cite{smith2019geometry} provide an overview of the role that different geometries play in continuous latent space models. For instance, spherical geometry, with its positive curvature, is well-suited for modeling networks with naturally bounded structures and homophilic clustering, where nodes are embedded on the surface of a sphere~\citep{mccormick2015latent}. In contrast, hyperbolic geometry provides a more parsimonious and effective framework for the hierarchical, tree-like structures common in complex networks. The volume of the hyperbolic space grows exponentially with its radius, which naturally accommodates the exponential growth of nodes around hubs in scale-free networks without requiring additional dimensions.

\cite{krioukov2010hyperbolic} introduced a physics-based model\footnote{We distinguish ``physics-based'' models---stochastic generative models designed to mimic real world phenomena---from (parametric) statistical models in which inference on unknown parameters and/or latent variables is prioritized.} for hyperbolic random graphs (HRG), mapping nodes onto a 2-dimensional hyperbolic disk. The connection probability between nodes follows a Fermi-Dirac distribution that depends on the hyperbolic distance between the nodes in the latent space and includes a temperature parameter $T$ that modulates the sharpness of this distance-dependence. Rooted in statistical physics, this hyperbolic framework provides a generative model with geometric intuition and has been used to study complex networks~\citep{boguna2010sustaining, voitalov2019scale, boguna2021network, osat2023embedding}. The translation of this physics model into a parametric statistical model remains an active area of research.

Statistical inference for hyperbolic network models is computationally challenging. Existing approaches include clique-based tests for latent space geometry ~\citep{lubold2023identifying}, maximum likelihood estimation via optimization on manifolds~\citep{kitsak2020link, li2026hyperbolic}, MCMC methods adapted to hyperbolic geometry~\citep{lizotte2025symmetry}, and variational methods~\citep{papamichalis2021latent}. All of these approaches fix Krioukov's temperature parameter at $T=0.5$ or omit it entirely. That is, they treat temperature as a fixed parameter. We argue that this simplification, which improves identifiability of other model parameters and is computationally convenient, sacrifices important model flexibility. In this work, we demonstrate that retaining the temperature parameter is crucial for capturing the true stochastic nature of network formation. We acknowledge that this creates a partially identified model~\citep{gustafson2010bayesian} but empirically show that the joint posterior of $(R, T)$, where $R$ is the hyperbolic disk radius, provides sufficient information for improved graph recovery performance. We develop two inference strategies: a fully Bayesian approach implemented in Stan~\citep{carpenter2017stan} for rigorous uncertainty quantification and an auto-encoding variational Bayes algorithm~\citep{kingma2013auto} for scalable inference on large networks.

The remainder of the paper is organized as follows. Section~\ref{sec:background} provides the necessary background on the continuous latent space model, with a particular focus on the hyperbolic geometry and the role of the temperature parameter in bridging the gap between hyperbolic and tree-like random graphs. In Section~\ref{sec:model}, we formally define our proposed model and the details of our two inferential procedures. Section~\ref{sec:experiment} validates our approach through a comprehensive simulation study, demonstrating the performance benefits of our model. Section~\ref{sec:discussion} concludes with a discussion of the implications of our findings and potential future research plans.

\section{Background}
\label{sec:background}

In this section, we introduce the continuous latent space (CLS) model framework and establish the mathematical foundations required for our proposed approach. CLS models assume that each node in an observed network occupies a position in an underlying continuous latent space, with the probability of edge formation between two nodes depending on their latent distance or similarity. This geometric approach to network modeling provides both interpretable low-dimensional representations and a principled framework for statistical inference. We emphasize how the choice of latent space geometry, across Euclidean, spherical, and hyperbolic spaces, fundamentally determines the model's capacity to represent different network structures (Section~\ref{sec:cls}). We then provide the mathematical background on hyperbolic geometry~\ref{sec:hyp_geom}. Finally, we examine the critical but often overlooked role of the temperature parameter in bridging the gap between idealized geometric structures and the stochastic nature of real-world networks in Section~\ref{sec:gap}.

\subsection{Continuous latent space models}
\label{sec:cls}
Following~\cite{smith2019geometry}, we consider the setting where the number of network nodes, $N$, indexed by $i, j\in\{1,..., N\}$, is fixed and where we observe the presence or absence of edges connecting pairs of nodes. We let $Y_{ij}$ be the random indicator of whether there is a tie (edge) between node $i$ and node $j$, $i\neq j$. We set $Y_{ii} =0$ to exclude self-loops and, for an undirected network, we assume $Y_{ij} = Y_{ji}$. Formally, a CLS model for the set of random variables $\mathbf{Y}=\{Y_{ij}: i,j=1,..., N\}$ is defined as follows:
 
\begin{definition}[Generic Continuous Latent Space Model]
\label{def:my_model}
Let $\mathbf{Y}=\{Y_{ij}\}$ be an undirected binary network on $N$ nodes. A continuous latent space model assumes
\begin{equation}
\label{eq:y_model}
Y_{ij} \sim \text{Bernoulli}(p_{ij}),\quad \text{where } p_{ij} = \sigma(s_{ij}|\mathbf{\gamma}).
\end{equation}
Here, $\sigma(\cdot)$ is a known, monotonically increasing link function parameterized by $\mathbf{\gamma}$, and $s_{ij}= s_{\mathcal{M}}(\mathbf{x}_i, \mathbf{x}_j)$ is the similarity between unknown ``position" parameters, $\textbf{x}_i$ and $\textbf{x}_j$. These positions $\{\textbf{x}_i\}$ lie on a $p$-dimensional Riemannian manifold $\mathcal{M}^p$, where $\textbf{x}_i\overset{\text{iid}}{\sim}f_{\mathcal{M}}(\theta)$  and $f_{\mathcal{M}}(\theta)$ is a distribution on $\mathcal{M}_p$ parameterized by $\theta$.
\end{definition}

The similarity measure $s_{\mathcal{M}}$ is a key modeling choice. For example,~\cite{mccormick2015latent} proposed a latent surface model using hyperspheres, where the greatest circle distance is a natural choice of similarity measure on a sphere. In many applications, however, the similarity $s_{ij}$ is defined as a (negative) distance function~\citep{hoff2002latent,smith2019geometry}, $s_{ij} = -d_{ij}$, so that nodes closer in the latent space have a higher probability of forming a tie. The model can be extended by including an intercept term, $\alpha$, to control the network's overall density, or by adding covariates to the link function.

CLS models embed nodes into a continuous geometric space, offering a natural means of capturing network phenomena. First, this geometric approach can account for transitivity, often described as the ``friend-of-a-friend" effect, whereby nodes that share connections with a mutual node are inferred to be nearby in the latent space. Additionally, CLS models incorporate latent homophily, the tendency for nearby nodes in the latent space to form ties. Both these features (transitivity and homophily) are commonly observed in real-world networks. The geometric framework also lends interpretability to the resulting embedding. By examining how nodes cluster or spread in the manifold, one may gain insights into community structure, outlier nodes, and other global or local network characteristics.

\begin{figure}[t]
    \centering
    \includegraphics[width=0.8\linewidth]{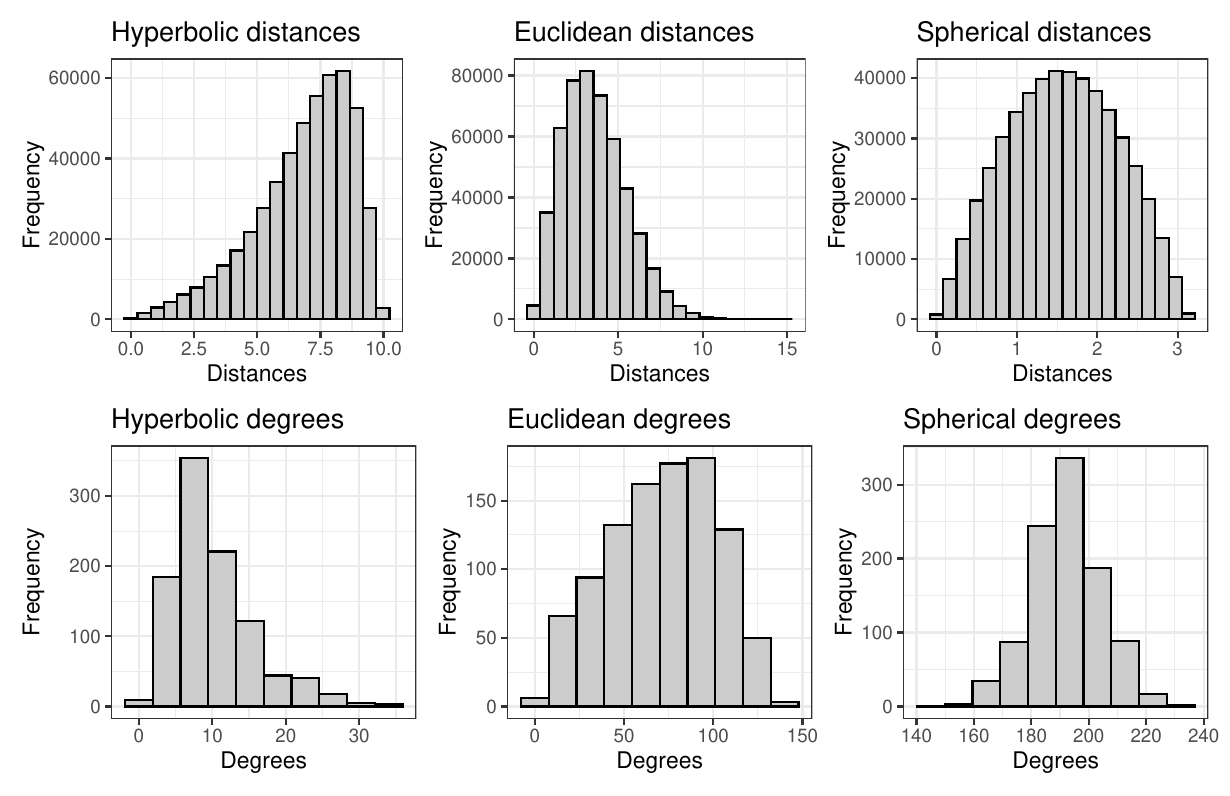}
    \caption{Histograms of distances in latent space and network degree distributions from networks generated from CLS with different latent space geometry (hyperbolic, Euclidean and spherical) with 1000 nodes, a logistic link function, and $\alpha=0$.}
    \label{fig:dist_deg_hist}
\end{figure}

Despite these appealing properties, applying CLS models presents significant practical challenges. A primary hurdle is the dimensionality of the parameter space. As the number of parameters scale linearly with the number of nodes, developing efficient inference procedures with accurate estimation becomes a challenge. The computational cost of estimating the latent positions can be prohibitive for large networks, motivating the need for efficient algorithms. A related challenge is the selection of an appropriate manifold. Similar to~\cite{smith2019geometry}, Figure~\ref{fig:dist_deg_hist} shows histograms of latent distances and resulting degree distributions for $N=1000$ nodes simulated under a CLS model with three latent geometries--- Euclidean, spherical, and hyperbolic--- using the same link function. Notably, the hyperbolic geometry yields a more power-law-like degree distribution, which aligns well with empirical observations in many large networks. The problem of selecting the appropriate geometry for the latent space has been explored in \cite{smith2019geometry} and \cite{lubold2023identifying}.  Here we instead focus on model specification for one particular geometry (hyperbolic) and computationally efficient inference for models within this class.   

In this paper we focus on CLS models with nodes embedded in hyperbolic manifolds with the similarity function defined to be the negative geodesic distance: $s_{ij}=-d_{ij}$, where $d_{ij} \equiv d(\mathbf{x}_i, \mathbf{x}_j)$ denotes the hyperbolic distance between $\mathbf{x}_i$ and $\mathbf{x}_j \in \mathbb{H}^p$. We use the Fermi-Dirac (logistic) link function to map distances to connection probabilities:

\begin{equation}
    \label{fermi-dirac}
    p_{ij} = \sigma(d_{ij}|\alpha, T) = \text{logistic}\left(\frac{\alpha-d_{ij}}{2T}\right)=  \frac{1}{1+exp(\frac{d_{ij}-\alpha}{2T})}.
\end{equation}
Additional model specification details needed for formal statistical inference are given in Section \ref{sec:model}.

While the H-CLS model is inspired by the  physics-based model of~\cite{krioukov2010hyperbolic}, our approach critically diverges from most statistical implementations of this CLS model due to the fact that we retain the temperature parameter $T$.  Here, $\alpha \in \mathbb{R}$ captures the overall density, and $T>0$ is the temperature parameter that controls the sharpness of the distance-probability relationship. When $T\longrightarrow 0$, the link function approaches a hard threshold at $\alpha$, so that the edges are formed almost deterministically according to whether the distances fall below this threshold. Similar distance-scaling parameters have been explored in other latent space contexts:~\cite{rastelli2016properties} include a scaling parameter $\phi$ in Euclidean CLS models with $f_\mathbb{R}(\theta)$ in Definition \ref{def:my_model} is the normal distribution. \cite{mccormick2015latent} introduce a parameter $\zeta$ for scaling angular distances on the hypersphere. In the hyperbolic CLS, formally defined below, we show later that this parameter $T$ plays a critical role in inducing tree-like structure. Following~\cite{krioukov2010hyperbolic}, we use ``tree-like" to refer to the hierarchical structure of the network, where nodes organize into groups and subgroups whose relationships can be approximated by a dendrogram, rather than the literal graph-theoretic absence of cycles. 

\subsection{Fundamentals of hyperbolic geometry}
\label{sec:hyp_geom}

Riemannian manifolds are smooth manifolds equipped with Riemannian metrics, inner products that vary smoothly on tangent spaces. An important property of a manifold is its curvature $\kappa$, which measures how much the space deviates from being flat (Euclidean space has zero curvature). Hyperbolic manifolds $\mathbb{H}_{\kappa}^p$ have negative curvature. An important geometric property of hyperbolic space is that geodesic distance increases exponentially with radial displacement from the origin. This results in volume growth that mirrors the branching pattern of trees, making hyperbolic space particularly well suited to modeling sparse, hierarchical, or tree-like networks.

Following~\cite{krioukov2010hyperbolic}, we work directly with polar coordinates in native hyperbolic space. Doing so simplifies computation and interpretation. For concreteness, we focus on $\mathbb{H}^2_{-1}$ (i.e., two-dimensional hyperbolic space with constant curvature $\kappa = -1$, though the framework extends to higher dimensions).
Rather than estimating curvature as a free parameter, we fix $\kappa=-1$ for hyperbolic latent space models, a convention in most theoretical and empirical work. This choice removes scaling ambiguity from the model and prevents further nonidentifiability when learning temperature and distances, which will be discussed in Section~\ref{sec:infer}.

We now provide definitions of the uniform distribution on a disc in $\mathbb{H}^2_{-1}$  and the geodesic distance between pairs of points on the disc. Let $\mathbb{H}_{-1}^2$ denote the unique non-compact, connected, Riemannian $2$-manifold with constant curvature -1, unique up to surjective isometry. We represent the points in $\mathbb{H}_{-1}^2$ as points in $\mathbb{R}^2$ given in polar coordinates $(r, \theta)$ in what is known as the \textit{native representation}. In this native representation, $(r,\theta)$ represents a point with intrinsic distance $r$ from $(0,0)$ to that point. 

\begin{definition}[Uniform distribution in hyperbolic space]
\label{def:unif}
 
A random point $(r,\theta) \in \mathbb{H}_{-1}^2$ is said to be uniformly distributed in a hyperbolic disk of radius $R$ centered at the origin $(0,0)$ if its density is proportional to the hyperbolic area measure.  We can generate such a point as follows: 
\begin{align*}
    \theta &\sim \text{Unif}(0,2\pi), \\
    u &\sim \text{Unif}(0,1), \\
    r &= \operatorname{arccosh}\!\bigl(1 + (\cosh R - 1)u\bigr).
\end{align*}
\end{definition}

This transformation for $r$ inverts the cumulative distribution induced by the hyperbolic area element, ensuring uniform point density despite exponential volume growth in $\mathbb{H}_{-1}^2$.

\begin{definition}[Hyperbolic distance in polar coordinates]
\label{def:hyp_dist}
For two points $(r_1,\theta_1)$ and $(r_2,\theta_2) \in \mathbb{H}_{-1}^2$, the hyperbolic geodesic distance is given by
\begin{align*}
    \Delta\theta &= \pi - \bigl|\pi - |\theta_1 - \theta_2|\bigr|, \\
    d\bigl((r_1,\theta_1),(r_2,\theta_2)\bigr)
    &= \operatorname{arccosh}\!\bigl(\cosh r_1 \cosh r_2
        - \sinh r_1 \sinh r_2 \cos \Delta\theta \bigr).
\end{align*}
\end{definition}

This polar coordinate parameterization of points in hyperbolic space facilitates interpretations of corresponding networks generated using the continuous latent space modeling framework. The radial coordinate $r$ captures node popularity or centrality (nodes with smaller $r$ are closer to the origin and tend to have higher degree), while the angular proximity between nodes reflects similarity in latent node attributes or community membership. This uniform distribution on hyperbolic space, combined with distance-based connection probabilities, generates the hierarchical, power-law structures characteristic of real networks and forms the basis of our approach.

\subsection{H-CLS models and tree-like networks}
\label{sec:gap}

Using Definitions \ref{def:unif} and \ref{def:hyp_dist}, we can formally define the hyperbolic continuous latent space model, which we refer to as the H-CLS model throughout the remainder of the paper. The H-CLS model is a special case of the model given by Definition \ref{def:my_model} in which the latent space is the $2$-dimensional hyperbolic manifold with constant constant curvature $\kappa = -1$, denoted $\mathbb{H}^2_{-1}$. Each node $i$ is associated with a latent position $\mathbf{x}_i=(r_i,\, \theta_i)$ drawn uniformly on hyperbolic disk of radius $R$ as described above. The similarity between nodes $i$ and $j$ is then defined as the negative geodesic distance, $s_{ij}=-d(\mathbf{x}_i, \, \mathbf{x}_j)=-d_{ij}$, where $d(\cdot,\,\cdot)$ is the hyperbolic distance in polar coordinates.

\begin{figure}
    \centering
    \begin{subfigure}[b]{0.48\textwidth}
        \centering
        \includegraphics[width=\linewidth]{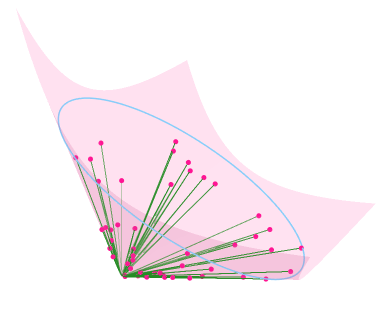}
        \caption{Hyperboloid model $\mathbb{H}_{-1}^2$.}
        \label{fig:hyp_vis}
    \end{subfigure}
    \hfill
    \begin{subfigure}[b]{0.48\textwidth}
        \centering
        \includegraphics[width=\linewidth]{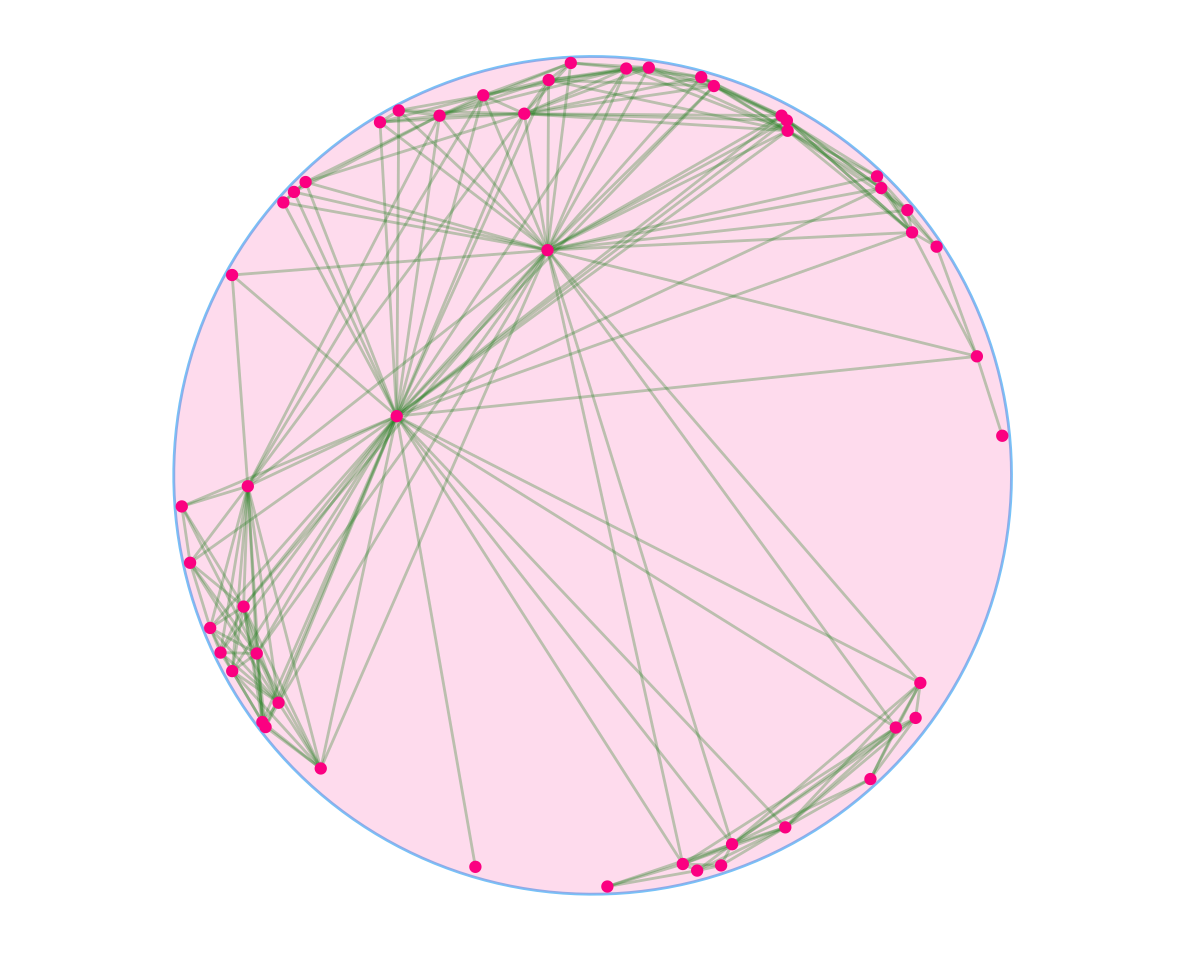}
        \caption{Poincaré disk model.}
        \label{fig:disk_vis}
    \end{subfigure}
    \caption{Visualizations using the hyperboloid model (left) and the Poincaré disk model (right) (see \citealp{smith2019geometry}) of a network embedded in a hyperbolic disk of radius $R=5$ (blue), with $\alpha=5$ and $T=0.01$.}
    \label{fig:hyp_both}
\end{figure}

As noted by \cite{krioukov2010hyperbolic}, in the H-CLS model, as $T\longrightarrow 0$, connections reduce to a hard threshold on hyperbolic distance. That is, viewed as a generative model, node connections become deterministic as $T\longrightarrow 0$:  $i$ and $j$ are connected if and only if $d_{ij}<\alpha$.  We can use this property of the small-temperature regime to motivate the connection between H-CLS models and tree-like networks. In $\mathbb{H}^2$, the circumference of a disc of radius $r$ grows approximately exponentially as $2\pi \sinh(r)$, so points distributed uniformly on the hyperbolic disc concentrate more at larger radii. Under the hard threshold, nodes at radius $r$ tend to connect to nodes at smaller radii. This radial connectivity pattern mirrors a tree hierarchy, with inner nodes acting as parents and outer nodes as children (see illustration in Figure~\ref{fig:hyp_both}). More generally, low temperature strengthens the geometric behavior necessary for tree-like topology, while high temperature relaxes these constraints, allowing non-geometric edges that create cycles and increase clustering.

\begin{figure}[ht!]
         \centering
         \includegraphics[width=0.9\textwidth]{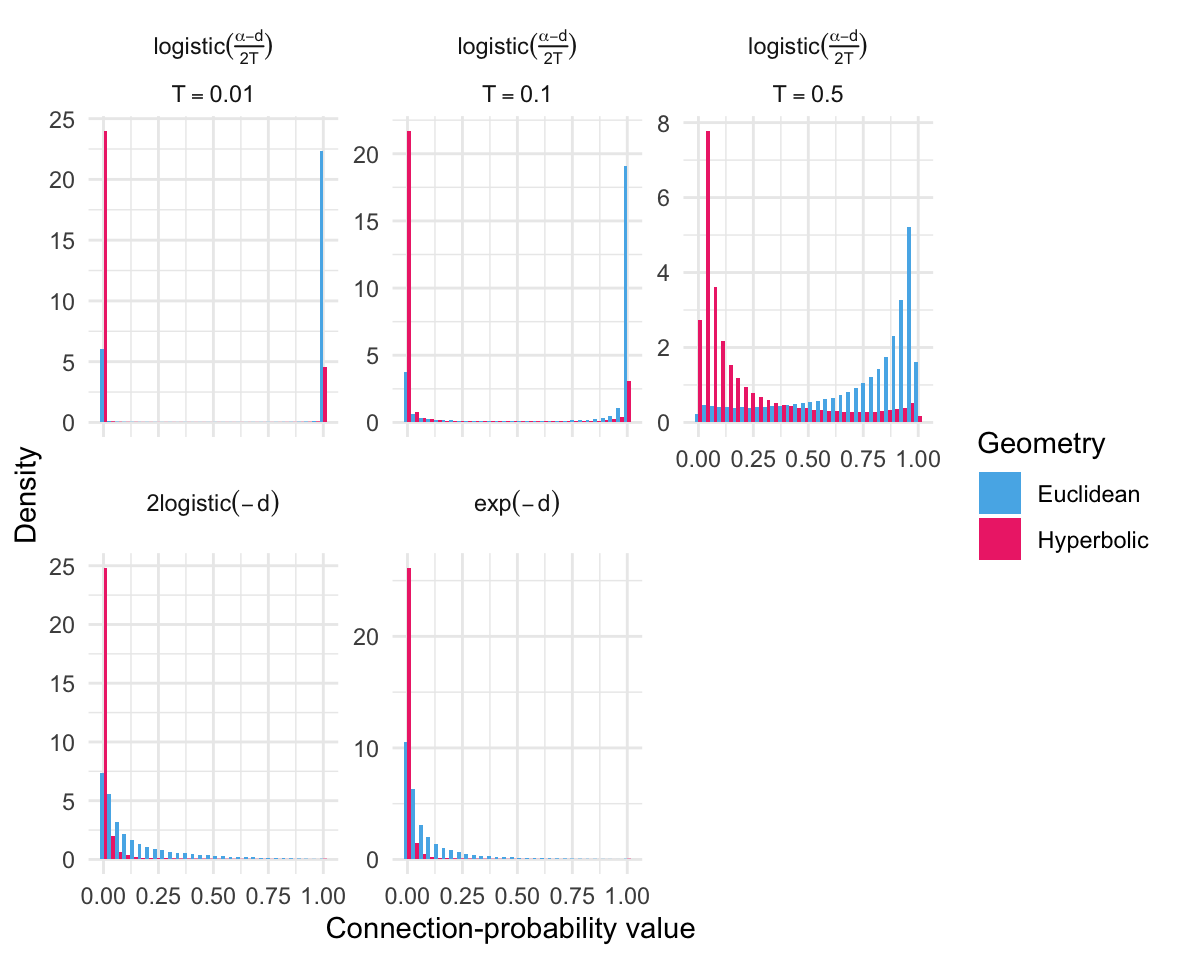}
        \caption{Connection probability distributions for the H-CLS and E-CLS models at different temperatures. 
        Histograms show the distribution of connection probabilities for networks generated with either hyperbolic (pink) or Euclidean (blue) geometry.}
        \label{fig:deg_link}
\end{figure}

We further explore the relationship between the functional form of the link function and networks generated under a generic CLS model with different underlying geometries. In particular, we simulate random graphs with $N=1000$ from H-CLS models with alternative link function and their Euclidean counterparts, E-CLSs. For the H-CLS model, nodes are positioned uniformly on the hyperbolic disc as in Definition \ref{def:unif} with fixed $R=5$. For the E-CLS model, latent positions are drawn from $N(0, \tau^2I_2)$, where $\tau = R/2.448$ is chosen so that 95\% of the Euclidean points fall within radius $R$, ensuring comparable spatial spread across geometries. Given the node positions, we compute pairwise distances $d_{ij}$ and apply various link functions $\sigma(d_{ij}|\alpha, T)$ to determine connection probabilities.  Figure~\ref{fig:deg_link} summarizes our findings. Each panel corresponds to a different link function.  The first three use Equation~\ref{fermi-dirac} at temperatures $T\in\{0.01, 0.1, 0.5\}$ and $\alpha=R$.  We also consider alternative forms that do not depend on the threshold $R$ or the temperature $T$ used in prior work: $2\text{logistic}(-d_{ij})$~\citep{li2026hyperbolic}, and $\text{exp}(-d_{ij})$~\citep{lubold2023identifying}.

Figure~\ref{fig:deg_link} reveals a clear contrast in the behavior of the H-CLS and E-CLS models across various link functions. At a low temperature ($T=0.01$ or $T=0.1$), the H-CLS model concentrates connection probabilities near 0, reflecting the fact that exponential volume growth in hyperbolic space causes most pairwise distances to be large. The E-CLS model, by contrast, concentrates probabilities near 1, as the matched spatial spread produces distances that largely fall within the connection threshold. As temperature increases to 0.5, the sharpness of the link function diminishes and both geometries produce more diffuse probability distributions, though their shapes remain approximately mirrored. The bottom row shows that link functions without a distance threshold or temperature concentrates probabilities near 0 with rapid decay for both geometries, indicating these link functions produce uniformly sparse connectivity.

\begin{figure}[ht!]
    \centering
    \begin{subfigure}[b]{0.3\textwidth}
        \centering
        \includegraphics[width=\textwidth]{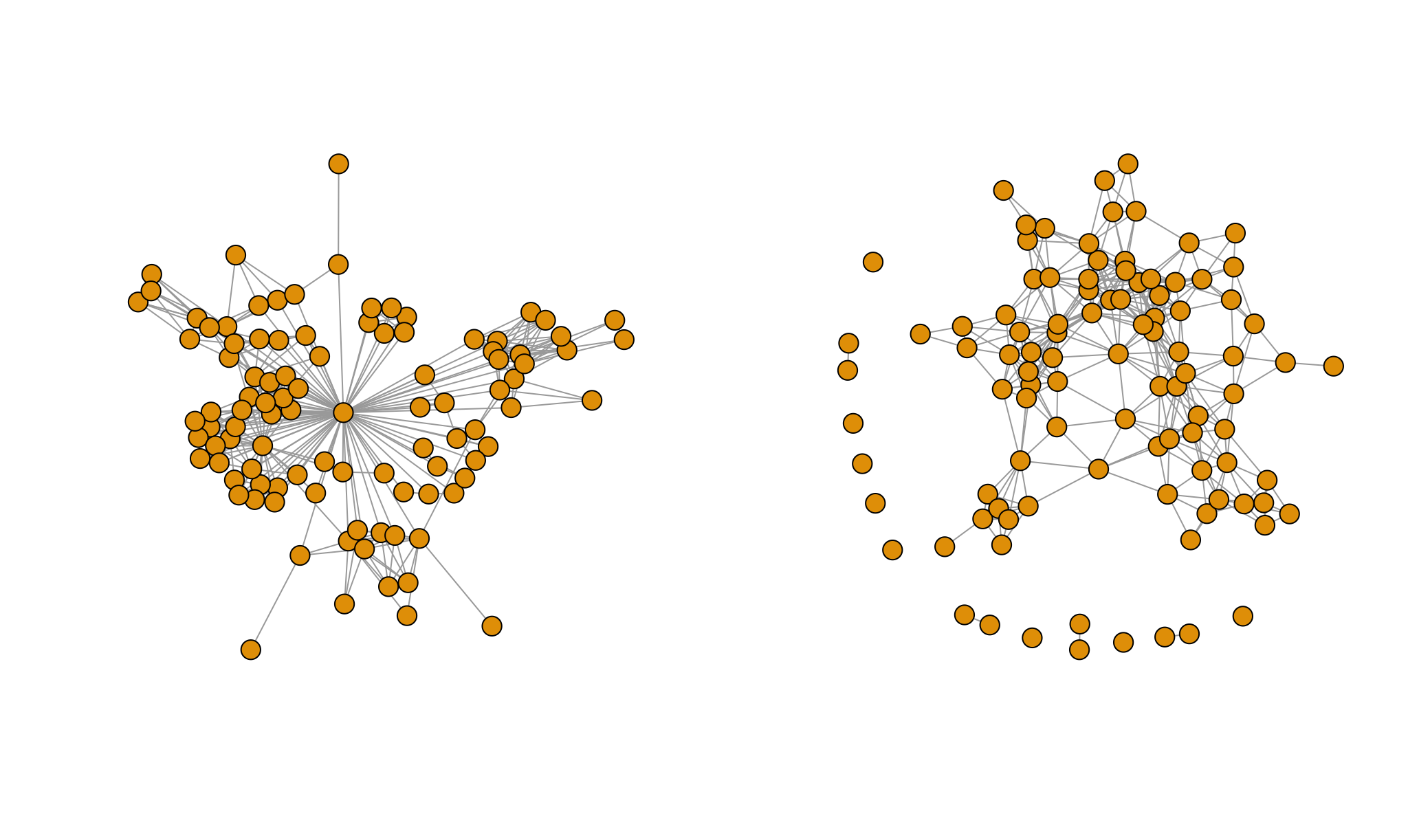}
        \caption{Left: H-CLS-generated network with $T=0.01$; Right: E-CLS-generated network with $T=0.01$.}
        \label{fig:hrg001b}
    \end{subfigure}
    \hfill
    \begin{subfigure}[b]{0.3\textwidth}
        \centering
        \includegraphics[width=\textwidth]{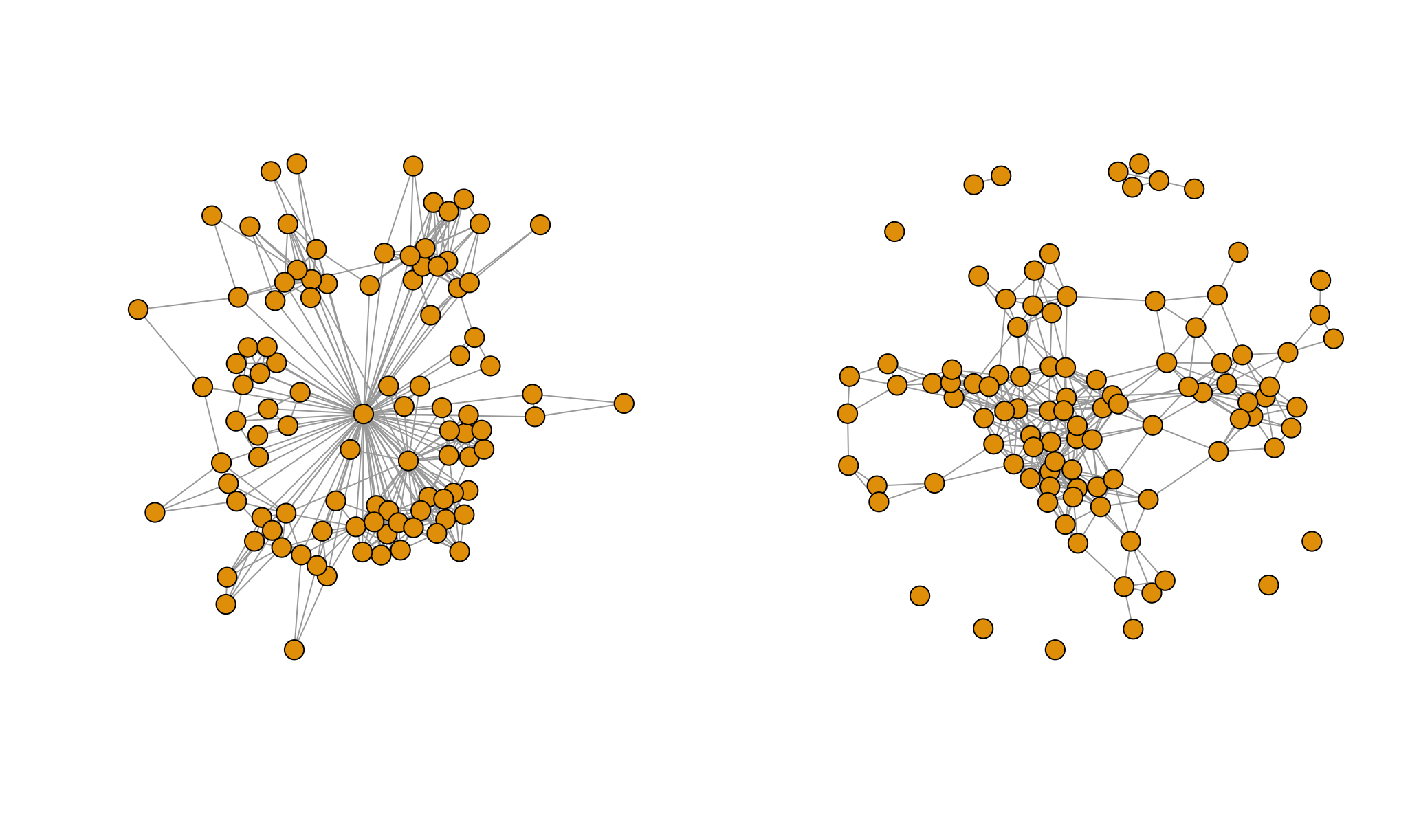}
        \caption{Left: H-CLS-generated network with $T=0.1$; Right: E-CLS-generated network with $T=0.1$.}
        \label{fig:hrg01b}
    \end{subfigure}
    \hfill
    \begin{subfigure}[b]{0.3\textwidth}
        \centering
        \includegraphics[width=\textwidth]{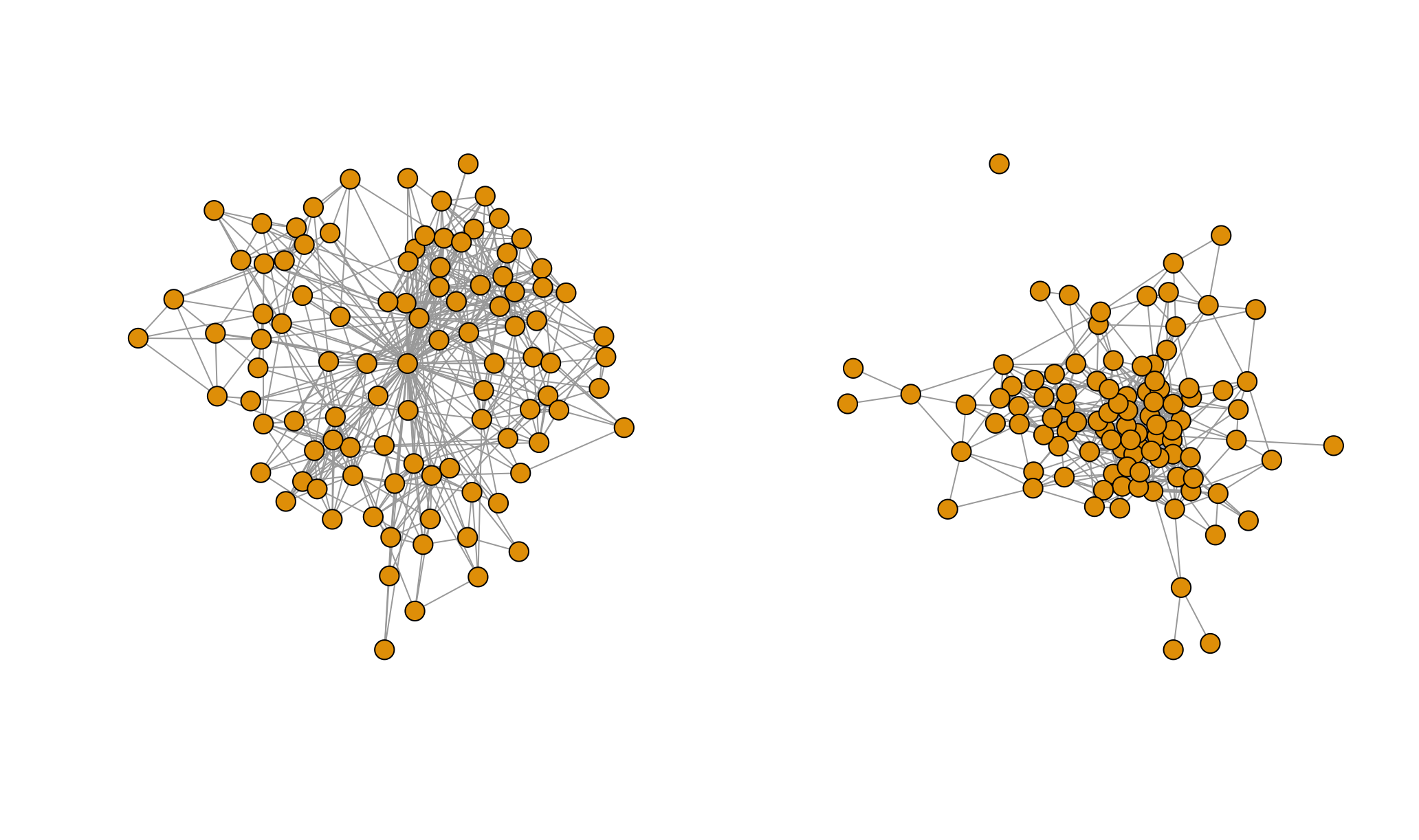}
        \caption{Left: H-CLS-generated network with $T=0.5$; Right: E-CLS-generated network with $T=0.5$.}
        \label{fig:hrg05b}
    \end{subfigure}

    \caption{Comparison of H-CLS- and E-CLS-generated networks under various temperatures.  Despite comparable sparsity, the H-CLS-generated networks consistently exhibit hierarchical structure, whereas E-CLS-generated networks produce more homogeneous connectivity typical of Euclidean space.}
    \label{fig:graph_vis}
\end{figure}

These differences in probability distributions are visually evident in the randomly-generated graphs shown in Figure~\ref{fig:graph_vis}. We generate random graphs with 100 nodes and radius $R=7$ from both the H-CLS and E-CLS models ($\tau=R/2.448$) under comparable conditions. Specifically, we adjust the $\alpha$ parameter in E-CLS model to match the edge density of the corresponding H-CLS, isolating geometry as the primary source of variation. Figure~\ref{fig:graph_vis} compares the H-CLS-generated graphs across three temperature levels $T\in\{0.01,0.1, 0.5\}$ against Euclidean counterparts. At low temperatures ($T=0.01$), connections are nearly deterministic based on latent distances, while higher temperatures introduce greater stochasticity, presenting more ``hairball-like'' topologies. Despite matched edge densities, the H-CLS exhibit clear tree-like structure dominated by central hubs, whereas the E-CLS lack prominent hub nodes. In addition, E-CLS models at low edge densities tend to be disconnected, particularly at low temperatures.

To demonstrate that this different behavior in these two geometries is a robust phenomenon and not an artifact of a single network realization, we then conducted a large-scale simulation study across 1000 replicates. We simulated networks from each geometry (Euclidean and hyperbolic) at temperature $T\in\{0.01, 0.1, 0.5\}$, with $N=100$ nodes and $\alpha=R=5$ ($\tau=R/2.448$ for the Euclidean points). The threshold $\alpha$ for the Euclidean space were calibrated as described above to produce similar edge densities across geometries. 

\begin{sloppypar}
While a rigorous quantification of tree-likeness would require measures such as $\delta$-hyperbolicity~\citep{chen2013hyperbolicity}, these are computationally prohibitive at scale. Instead, we use a panel of structural metrics as tractable proxies, motivated by the properties of trees: a perfect tree has zero circuit rank, zero clustering coefficient, and maximal path lengths relative to its size. We compute the following summary statistics for each realization. Let $G=(V, E)$ denote a graph with $|V|=N$ nodes and $|E|=m$ edges, and let $d(i, j)$ denote the shortest path length between nodes $i$ and $j$:
\end{sloppypar}
\begin{itemize}
    \item Circuit rank: $c(G) = m-N+k$, where $k$ is the number of connected components; trees have $c=0$, so this directly measures how far $G$ is from being a tree.
    \item Global clustering coefficient: $C(G)=\frac{3\times \text{number of triangles}}{\text{number of connected triples}}$; trees have no triangles, so low clustering suggests tree-like structure.
    \item Mean betweenness centrality (normalized): $\bar{B}(G)=\frac{1}{N}\sum_{i\in V}\frac{\sum_{s\neq i\neq t}\lambda_{st}(i)/\lambda_{st}}{(N-1)(N-2)}$, where $\lambda_{st}$ is the number of shortest paths from $s$ to $t$ and $\lambda_{st}(i)$ is the number passing through $i$; trees have high betweenness concentrated on internal nodes.
    \item Mean closeness centrality (normalized): $\bar{L}(G)=\frac{1}{N}\sum_{i\in V}\frac{N-1}{\sum_{j\neq i}d(i,j)}$; trees lack shortcuts, so closeness is lower overall.
    \item Mean eigenvector centrality: $\bar{E}(G)=\frac{1}{N}\sum_{i\in V} z_i$, where $\mathbf{z}$ is the leading eigenvector of the adjacency matrix; in trees, influence concentrates at a few hub nodes while most leaves contribute little, resulting in low mean.
    \item Mean path length: $l(G) = \frac{1}{N(N-1)}\sum_{i\neq j} d(i,j)$; trees tend to have longer paths than densely connected graphs.
    \item Modularity: $Q(C)=\frac{1}{2m}\sum_{ij}[Y_{ij}-\frac{k_ik_j}{2m}\delta(c_i, c_j)]$, where $c_i$ denotes the community assignment of node $i$, obtained using the fast greedy modularity optimizing algorithm~\citep{clauset2004finding}; hierarchical networks naturally decompose into nested communities, yielding high modularity. 
\end{itemize}

\begin{figure}[ht!]
         \centering
         \includegraphics[width=\textwidth]{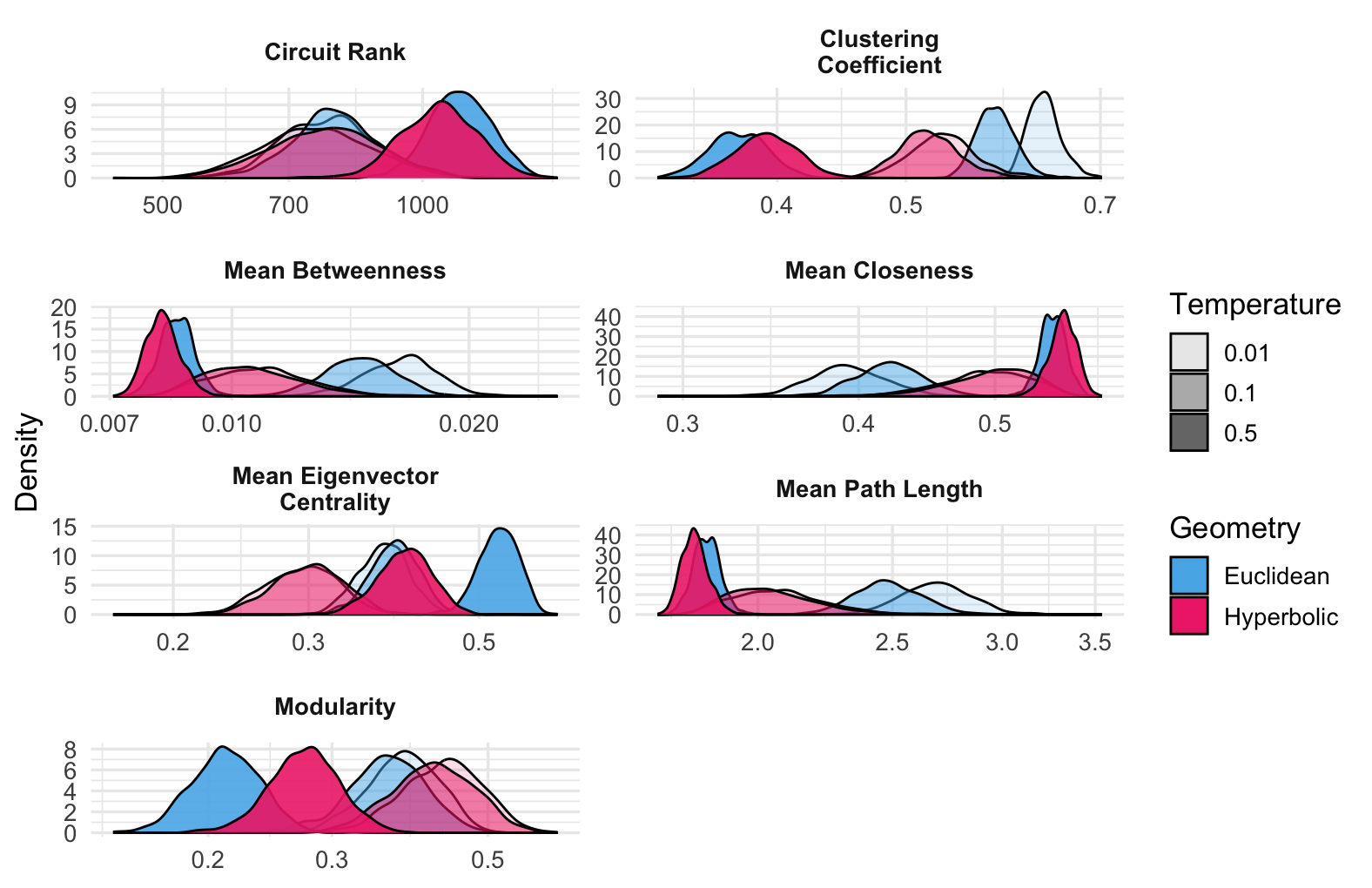}
        \caption{Empirical density distributions of circuit rank (top left panel), clustering coefficient (top right panel), mean betweenness (normalized, second row left), mean closeness (normalized, second row right), mean eigenvector centrality (third row left), mean path length (third row right) and modularity (log-scale) for 1000 network generated under different geometries with $N$ = 100 nodes and $R$ = 5. $\alpha$ is adjusted so that the edge densities are similar across geometries (hyperbolic and Euclidean) and temperatures ($T\in\{0.01, 0.1, 0.5\}$), shaded from light to dark. At high temperature, both geometries produce largely overlapping distributions, suggesting limited structural differentiation. Lower temperatures yield higher clustering and lower circuit rank for both models.}
        \label{fig:treelikeness}
\end{figure}

Figure~\ref{fig:treelikeness} summarizes these results, showing that hyperbolic networks at low temperature exhibit structural signatures most consistent with tree-like organization. No single metric perfectly captures tree-likeness; for instance, the clustering coefficient increases at low temperature despite trees being triangle-free, reflecting that geometric transitivity in the latent space generates local triangles even as the global structure becomes more hierarchical. However, taken together, these metrics create a consistent picture. 

Both geometries exhibit sensitivity to temperature: as $T$ decreases, the distributions sharpen and shift. At the conventional temperature $T=0.5$, the distributions for the H-CLS and E-CLS models largely overlap across most of the metrics, with the exceptions of modularity and mean eigenvector centrality, which retain clear separation between the two geometries even at high temperatures. More broadly, the two geometries become increasingly distinguishable as temperature decreases, suggesting the geometric effects on network structure are stronger in the low-temperature regime.

This implies that while hyperbolic geometry provides the necessary foundation for sparse hierarchical graphs, it is the low-temperature regime that enforces tree-like structure. At high temperature, the hyperbolic model produces networks that are difficult to distinguish from Euclidean networks based on these summary statistics. The distinctive hierarchical properties often attributed to hyperbolic geometry, high clustering and tree-like organization, emerge primarily at low temperatures. This suggests that fixing $T=0.5$, as is conventional in many implementations, may obscure the structural features that motivate the use of hyperbolic latent space models.

These results demonstrate that temperature is not merely a parameter to fix for convenience, but fundamentally controls the tree-like structure of hyperbolic random graphs and their ability to model the hierarchical properties of real-world networks. Therefore, while existing statistical implementations fix $T$ or omit it entirely, this choice sacrifices the model's ability to capture the varying degrees of tree-likeness in real networks. Our findings provide empirical support for estimating the temperature from data. Doing so allows the model to adapt to the specific structural properties of a network, which may yield a more accurate and robust latent space representation.  

In related work, \cite{li2026hyperbolic} proposed a hyperbolic network latent space model with learnable curvature, demonstrating both theoretically and empirically that fixing the curvature to a default value can lead to suboptimal embeddings. While their work also addresses hyperbolic network modeling, their formulation adopts a generic logistic link function of hyperbolic distance ($2\text{logistic}(-d_{ij})$), dropping both the temperature $T$ and radius $R$, which are central to the model of~\cite{krioukov2010hyperbolic}. Without specifying the radius $R$ as in Definition~\ref{fermi-dirac}, the curvature $\kappa$ in their work is forced to simultaneously capture the intrinsic geometry of the latent space and control the overall network sparsity, a role that $R$ would otherwise play. This conflation may raise concerns about the interpretability of the resulting embeddings, as the estimated curvature and latent positions may reflect density corrections rather than actual latent structure.

\section{Complete H-CLS model specification and inference}
\label{sec:model}

The following subsections complete the specification of the H-CLS model introduced in Section \ref{sec:gap} and present two posterior inference strategies: a full Bayesian approach via Hamiltonian Monte Carlo (HMC) and a scalable variational inference algorithm.

\subsection{Prior specification}
\label{sec:prior}

The full Bayesian specification of the H-CLS model includes priors on the latent node positions and on the global parameters including the disk radius $R$, the connection threshold $\alpha$, and the temperature $T$. 
The prior for the latent positions, $\mathbf{x}_i$, assumes nodes are distributed uniformly within a hyperbolic disk of radius $R$ as described in Section~\ref{sec:hyp_geom}. In our model, this radius $R$ is not a fixed hyperparameter but is instead inferred from the data, allowing the data to inform the appropriate geometric scale of the network. We put a weakly informative prior on $R$ with $R\sim \text{Exponential}(1)$. We further specify $R$ to govern both the size of the latent space and the mean of the connection threshold $\alpha$, coupling the geometric scale to the model's sparsity. We place an informative prior $\alpha\sim \text{Normal}(R, 0.1)$ centered at the global scale $R$. The tight prior concentrates $\alpha$ near $R$ with high probability, while permitting deviations that yield a soft geometric boundary rather than a hard radial cutoff. This approach allows the model to be more adaptable to the noisy, non-ideal nature of real-world networks.
The temperature parameter, $T$, upper-bounded at 0.5 controls the sharpness of the distance-to-probability mapping and governs the network's tree-likeness (Section~\ref{sec:gap}). We place a weakly informative $\text{Gamma}(0.1,1)$ prior on $T$ favoring the low-temperature regimes consistent with sparse, hierarchical structures, while still allowing the data to support higher temperatures if necessary.

As is common in latent space models, the individual latent positions $\mathbf{X}=\{\mathbf{x}_i\}$ are not strictly identifiable due to the rotational and reflectional invariance of hyperbolic space. However, our inferential target is the set of pairwise geodesic distances $d_{ij}$ and the global model parameters, which are invariant under these isometries, and are sufficient for tasks such as graph recovery and link prediction.


As $\kappa$ and $T$ are not separately identifiable (see Section~\ref{sec:background}), we fix $\kappa = -1$.  The identifiability considerations among the remaining parameters are discussed in Section~\ref{sec:infer}.

Inference for the model proceeds via the posterior distribution of all unknown quantities given the observed network $\mathbf{Y}$. The posterior is given by:
\begin{equation}
\label{eqn:posterior}
    p(\mathbf{X},R,\alpha,T|\mathbf{Y})\propto p(\mathbf{Y}|\mathbf{X},\alpha,T)p(\mathbf{X}|R)p(\alpha|R)p(R)p(T).
\end{equation}

\subsection{Inference}
\label{sec:infer}
This section details the computational strategies for fitting the \Mymodel{} model defined earlier. We present two inferential approaches tailored to networks of different size. For smaller networks, where a full posterior characterization is feasible, we use HMC for rigorous uncertainty quantification. For larger networks, HMC becomes computationally prohibitive so we propose a scalable variational inference algorithm. 

\subsubsection{Full posterior}
For rigorous posterior inference and full uncertainty quantification, we fitted the \Mymodel{} model using HMC by leveraging the Stan probabilistic programming language~\citep{stan2024} and the functionality of the PyMC3 Python package~\citep{salvatier2016probabilistic}. We found that this inferential strategy requires careful handling of the hyperbolic geometry. Specifically, we parameterized the node positions, $\mathbf{x}_i$, using polar coordinates, $(\theta_i, r_i)$. To make the HMC algorithm more numerically stable, we rewrite the pairwise distance as $$d_{ij} = arccosh\left(1+2\left(\sinh^2\left(\frac{r_i-r_j}{2}\right)+\sinh r_i\sinh{r_j}\sin^2\frac{\Delta \theta_{ij}}{2}\right)\right),$$
where $\Delta \theta_{ij} = \vert \theta_i - \theta_j\vert$.  This reparametrization avoids numerical overflow for large radial coordinates. In addition, we use the log-odds parameterization in the Stan sampling function \textit{bernoulli\_logit} to further encourage numerical stability. 

\begin{figure}[!ht]
    \centering
    \begin{subfigure}{0.9\textwidth}
        \centering
        \includegraphics[width=0.48\textwidth]{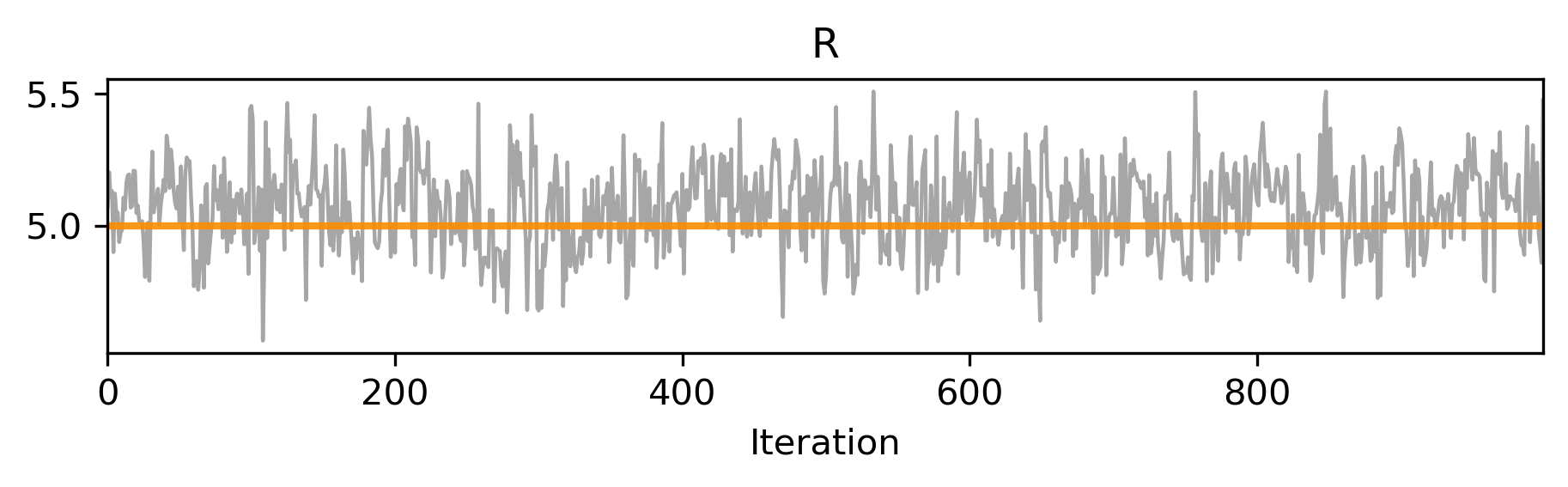}\hfill
        \includegraphics[width=0.48\textwidth]{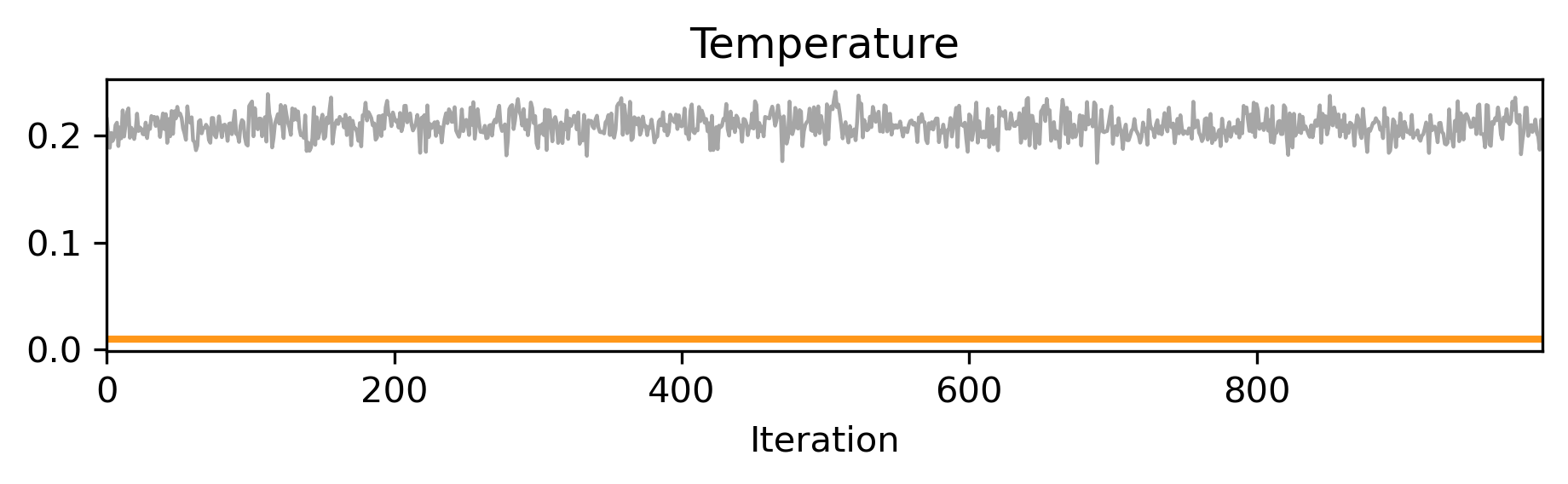}
        
        \vspace{1em}
        
        \includegraphics[width=0.48\textwidth]{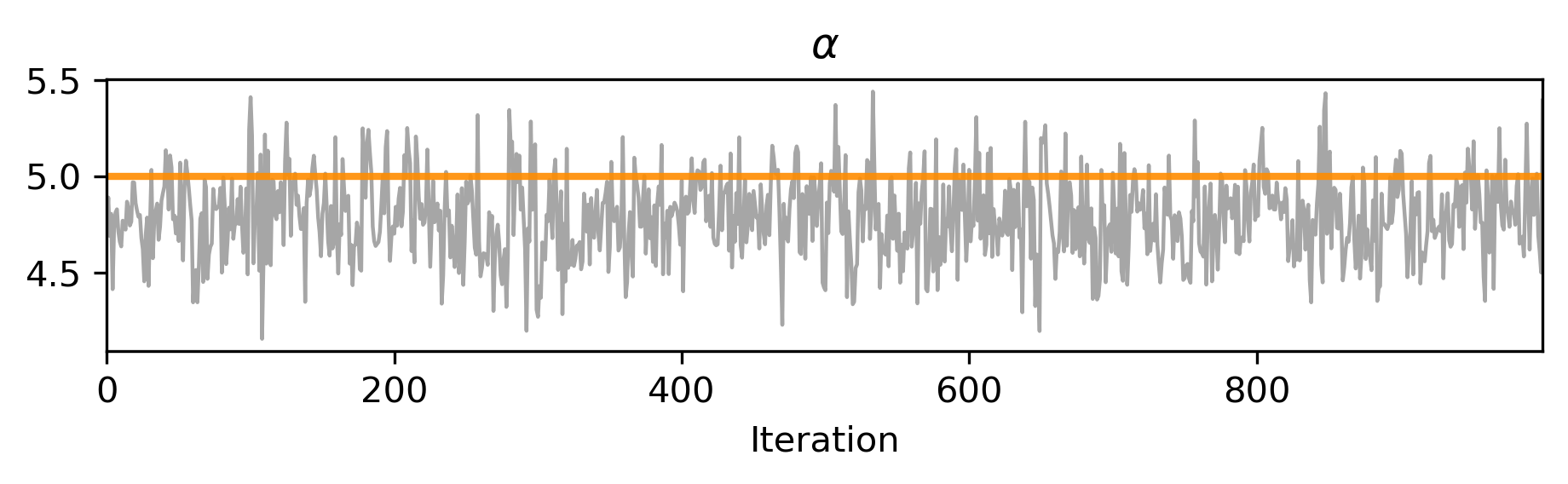}\hfill
        \includegraphics[width=0.48\textwidth]{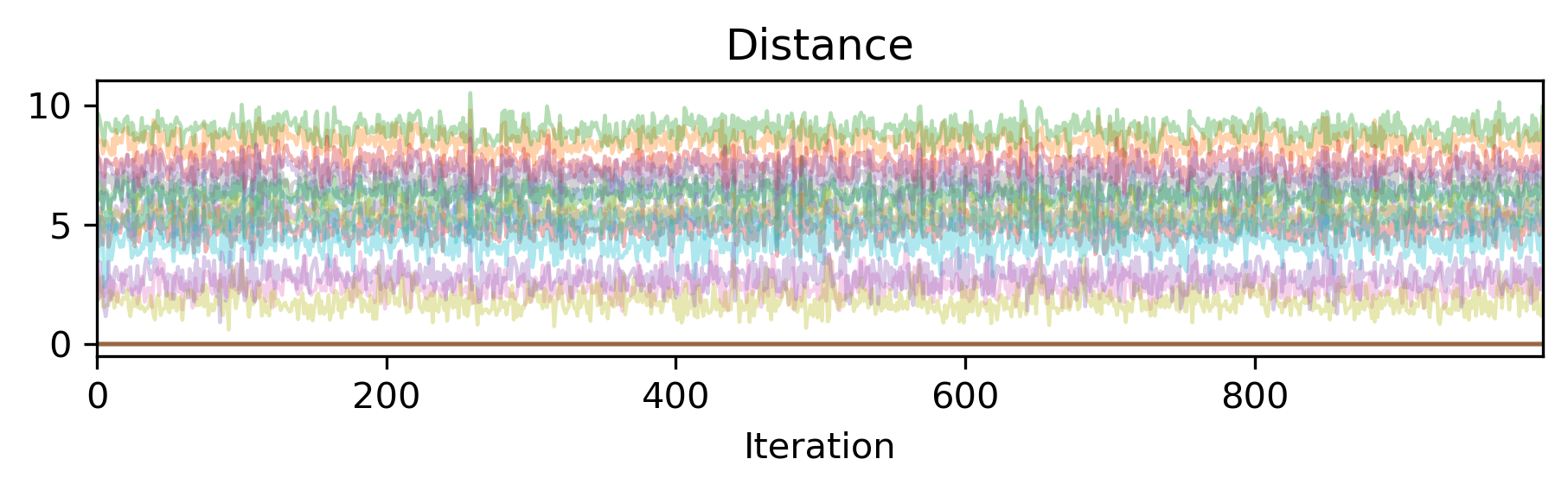}
        
        \label{fig:mcmc_traces}
    \end{subfigure}

    \caption{MCMC diagnostics and performance on a representative simulated network (N=100). Orange horizontal lines indicate true values. Traceplots demonstrate good mixing and convergence.} 
    \label{fig:mcmc_validation_combined}
\end{figure}

To validate the Stan HMC implementation, we fit the model to data generated from the model with $N=100$ nodes, a true radius $R=5$, and a true temperature $T=0.01$. In Figure~\ref{fig:mcmc_validation_combined} we present the traceplots for $R$, $\alpha$, temperature, and some pairwise distances. We can see evidence that the Stan HMC mixes well. As mentioned earlier, we do not expect full identifiability from this model, so we can see the sample path of the temperature parameter does not cover its true value. 

To measure model fit, we calculated the area under the curve (AUC), the integral of the true positive rate as a function of the false positive rate over all classification thresholds. 
For our simulated data, the H-CLS model fitted using HMC achieves a 0.984 AUC in graph reconstruction when fitted to data generated from the correct model, $3\%$ higher than the AUC obtained when fitting the incorrect model with temperature fixed at 0.5. Additionally, strongly significant correlations were observed between true, unobserved distances and the inferred distances (Pearson $r = 0.946$, Spearman $\rho = 0.943$, both $p < 0.001$), while the fixed temperature model results in lower correlations (Pearson $r=0.820$ and Spearman $\rho=0.740$). Results of a more extensive simulation study are provided in Section~\ref{sec:experiment}.

While the HMC approach provides full posterior inference, its computational complexity of $O(N^2)$, limiting its practical application to networks with fewer than a few hundreds of nodes, motivating our variational alternative for larger networks. 

\subsubsection{Variational inference}
\label{sec:vi}

For scalable inference on large networks, we develop an auto-encoding variational Bayes (AEVB) algorithm. The central challenge is that the posterior over the latent hyperbolic coordinates is intractable, and defining a variational family directly on the hyperbolic disk is problematic: the Kullback-Leibler (KL) divergence between the variational distribution and the uniform hyperbolic prior lacks a closed-form expression, and numerical approximations can be unstable. Our strategy, therefore, is to perform variational inference in an intermediate Euclidean space and then map the result to the hyperbolic disk through a fixed, deterministic transformation.  
Specifically, for each node $i$, we introduce two auxiliary variables $(z_{i, r}, z_{i, \theta})\in \mathbb{R}^2$ and define the approximate posterior as a fully factorized Gaussian,
$$q_{\phi}(z_{i, r}|\mathbf{Y})=N(\mu_{i, r}, \sigma_{i,r}^2), \quad q_{\phi}(z_{i, \theta}|\mathbf{Y}) = N(\mu_{i, \theta}, \sigma^2_{i, \theta}),$$ 
whose parameters $\phi=\{\mu_{i, r},\mu_{i,\theta},\sigma^2_{i,r},\sigma^2_{i,\theta}\}_{i=1}^N$ are output by an encoder network. The encoder is a two-layer graph convolutional network (GCN) that takes the adjacency matrix $\mathbf{Y}$ as input to produce node-level representations, enabling amortized inference across all nodes simultaneously. Given a sample $(z_{i,r}, z_{i,\theta})\sim q_{\phi}$, the hyperbolic coordinates $(r_i,\theta_i)$ are obtained by the deterministic mapping
\begin{align*}
    u_i &= sigmoid(z_{i,r}) \\
    r_i &= arccosh(1+(\cosh R-1)\cdot u_i) \\
    \theta_i &= z_{i, \theta} \pmod{2\pi}.
\end{align*}
By placing a Gaussian prior on $z_{i,r}$, the transformed variable $u_i$ follows a logit-normal distribution on $(0,1)$, which provides a more flexible model for the radial coordinate than the uniform hyperbolic prior assumed by our H-CLS model.  

Inference proceeds by maximizing the Evidence Lower Bound (ELBO) with respect to the variational parameters $\phi$ and the global model parameters $\Phi=\{R, \alpha, T\}$. The objective function is 
$$\mathcal{L}(\phi, \Phi) = \mathbb{E}_{q_{\phi}}[\log p(\mathbf{Y}|Z, \Phi)] - D_{KL} (q_{\phi} (Z|A)||p(Z))+\log p(\Phi),$$
where $Z=\{(z_{i, r}, z_{i,\theta})\}_{i=1}^N$ denotes the full set of Euclidean auxiliary variables, $p(Y|Z,\Phi)$ is evaluated with the Bernoulli likelihood, and the KL divergence has an analytical form $$D_{KL}(q_{\phi}(Z|A)||p(Z))=-0.5  \sum_i(1+\log \sigma_{r,i}^2-\mu_{r,i}^2-\sigma_{r,i}^2)-0.5\sum_i(1+\log \sigma_{\theta,i}^2-\mu_{\theta,i}^2-\sigma_{\theta_i}^2),$$
where $\sigma_{r,i}$, $\mu_{r,i}$, $\sigma_{\theta,i}$, and $\mu_{\theta,i}$ are all outputs of the encoders after the transformation described above.

Crucially, the KL divergence term is computed in the Euclidean latent space between the variational Gaussians and standard normal priors, which has a closed-form solution, ensuring stable optimization. The expectation over the likelihood is approximated via the reparameterization trick as in~\cite{kingma2013auto}. This formulation scales to networks with thousands of nodes, where MCMC methods become prohibitive. 

\section{Simulation studies, sensitivity analyses, and real data examples}
\label{sec:experiment}
In this section, we evaluate the \Mymodel{} model through a comprehensive series of simulation studies designed to assess its performance. We conduct a large-scale simulation study generating networks from the H-CLS model with low temperature varying network size, $N$, and radii, $R$, to systematically compare the performance of alternative CLS mdoels. To assess model robustness under misspecification, we further simulate networks from CLS with a different underlying geometies, Euclidean and hyperbolic with a fixed temperature parameter, allowing us to quantify the costs of geometric and parametric misspecification. Finally, we evaluate the scalability of our variational inference framework on significantly larger networks, demonstrating the model's applicability to real-world networks where full MCMC inference may be computationally expensive. 

As described above, we evaluate model performance through a graph reconstruction task. For each simulated network, the model is fitted on the full adjacency matrix $\mathbf{Y}=\{Y_{ij}\}$ with the goal of learning a latent representation that accurately reproduces the observed graph structure. We assess the model's in-sample fit by comparing its predicted links $\hat{y}_{ij}$ to the ground-truth edges $\mathbf{Y}_{ij}$. Performance is measured by AUC, where again higher values indicate better reconstruction,  and by overall reconstruction accuracy (the proportion that the prediction matches the observed binary outcome).

\subsection{Model performance comparison}

We first simulate 30 networks from the H-CLS model with $T=0.01$ for each combination of $N\in \{30, 50, 100, 150, 200\}$, $R\in \{3, 5, 7, 10\}$ and $\alpha=R$. We fit three candidate models to each simulated graph: a hyperbolic model with learnable temperature (H-CLS), a hyperbolic model with temperature fixed at 0.5 (H-CLS (T=0.5)), and a two-dimensional E-CLS model. The H-CLS (T=0.5) model is similar to the H-CLS model, where we fix the temperature at 0.5 instead of estimating it. The E-CLS model assumes the latent positions follow a bivariate normal distribution on $\mathbb{R}^2$ centered at zero: $\mathbf{x}_i\sim \text{Normal}(0, \tau^2)$, where $\tau$ is to be estimated with a $\text{Gamma}(1,1)$ prior.

In Table~\ref{tab:model_comparison}, we compare performance among the three models by listing the average and maximum AUC/accuracy, across all simulated networks, as well as the proportion of simulations in which each model outperforms the others. Across simulation settings, no single model uniformly dominates. The hyperbolic model with estimable temperature performs competitively throughout, with the highest or tied-highest proportion of correctly predicted edges in the majority of settings. This advantage becomes more pronounced as the network size increases: at $N=200$, the estimable-temperature model dominates across nearly all metrics for $R \leq 7$.

Despite its flexibility, the estimable-temperature model does not uniformly outperform the simpler alternative, the Euclidean model. This reflects a bias-variance tradeoff: increased model complexity introduces estimation uncertainty that can offset gains from correct specification, particularly when network sizes are small. Additionally, small networks relative to their latent radius may not contain sufficient information to reveal the hierarchical structure inherent in hyperbolic geometry; as network size increases, the structure becomes more apparent and the H-CLS model shows clearer advantages. In contrast, the fixed-temperature hyperbolic model rarely achieves the best performance, despite assuming the correct underlying geometry. This suggests that fixing the temperature at an incorrect value is more detrimental than  misspecification of the latent geometry: the Euclidean model, while assuming the wrong geometry, outperforms the fixed-temperature hyperbolic model in most settings.

\begin{table}[htbp]
\centering
\caption{Model comparison across network sizes ($N$) and radius values ($R$). Bold indicates highest value among models for each metric within each $(N, R)$ combination.}
\label{tab:model_comparison}
\resizebox{\textwidth}{!}{%
\begin{tabular}{llr|cc|cc|cc|cc}
\hline
& & & \multicolumn{2}{c|}{R = 3.0} & \multicolumn{2}{c|}{R = 5.0} & \multicolumn{2}{c|}{R = 7.0} & \multicolumn{2}{c}{R = 10.0} \\
N & Model & Metric & AUC & Accuracy & AUC & Accuracy & AUC & Accuracy & AUC & Accuracy \\
\hline
\multirow{9}{*}{30} 
& \multirow{3}{*}{E-CLS} 
  & Average & \textbf{0.993} & 0.903 & 0.986 & 0.892 & \textbf{0.997} & \textbf{0.943} & \textbf{1.000} & 0.982 \\
& & Max & \textbf{1.000} & \textbf{0.996} & \textbf{1.000} & 0.996 & \textbf{1.000} & \textbf{1.000} & \textbf{1.000} & \textbf{1.000} \\
& & Prop. Best & \textbf{0.567} & 0.433 & \textbf{0.600} & 0.367 & \textbf{0.833} & \textbf{0.500} & \textbf{0.567} & 0.233 \\
\cline{2-11}
& \multirow{3}{*}{H-CLS (T=0.5)} 
  & Average & 0.970 & 0.796 & 0.957 & 0.875 & 0.938 & 0.935 & 0.945 & 0.980 \\
& & Max & 0.993 & 0.840 & 0.992 & 0.929 & 0.983 & 0.962 & \textbf{1.000} & 0.994 \\
& & Prop. Best & 0.033 & 0.000 & 0.033 & 0.200 & 0.000 & 0.333 & 0.000 & 0.167 \\
\cline{2-11}
& \multirow{3}{*}{H-CLS} 
  & Average & 0.989 & \textbf{0.912} & \textbf{0.988} & \textbf{0.920} & 0.988 & 0.936 & 0.991 & \textbf{0.985} \\
& & Max & \textbf{1.000} & 0.994 & \textbf{1.000} & \textbf{1.000} & \textbf{1.000} & \textbf{1.000} & \textbf{1.000} & \textbf{1.000} \\
& & Prop. Best & 0.400 & \textbf{0.567} & 0.367 & \textbf{0.433} & 0.167 & 0.167 & 0.433 & \textbf{0.600} \\
\hline
\multirow{9}{*}{50} 
& \multirow{3}{*}{E-CLS} 
  & Average & \textbf{0.985} & 0.887 & 0.971 & 0.884 & \textbf{0.990} & \textbf{0.946} & \textbf{1.000} & \textbf{0.988} \\
& & Max & 0.998 & 0.936 & 0.996 & 0.943 & \textbf{1.000} & \textbf{0.998} & \textbf{1.000} & \textbf{1.000} \\
& & Prop. Best & \textbf{0.500} & 0.433 & \textbf{0.467} & 0.300 & \textbf{0.867} & \textbf{0.600} & \textbf{0.567} & \textbf{0.467} \\
\cline{2-11}
& \multirow{3}{*}{H-CLS (T=0.5)} 
  & Average & 0.966 & 0.803 & 0.954 & 0.879 & 0.924 & 0.937 & 0.915 & 0.980 \\
& & Max & 0.988 & 0.827 & 0.985 & 0.906 & 0.974 & 0.962 & 0.986 & 0.990 \\
& & Prop. Best & 0.100 & 0.000 & 0.100 & 0.200 & 0.033 & 0.300 & 0.000 & 0.233 \\
\cline{2-11}
& \multirow{3}{*}{H-CLS} 
  & Average & 0.983 & \textbf{0.902} & \textbf{0.982} & \textbf{0.915} & 0.981 & 0.943 & 0.974 & 0.985 \\
& & Max & \textbf{1.000} & \textbf{0.997} & \textbf{0.999} & \textbf{0.967} & \textbf{1.000} & 0.987 & \textbf{1.000} & \textbf{1.000} \\
& & Prop. Best & 0.400 & \textbf{0.567} & 0.433 & \textbf{0.500} & 0.100 & 0.100 & 0.433 & 0.300 \\
\hline
\multirow{9}{*}{100} 
& \multirow{3}{*}{E-CLS} 
  & Average & 0.980 & 0.882 & 0.960 & 0.876 & \textbf{0.972} & \textbf{0.937} & \textbf{0.999} & \textbf{0.985} \\
& & Max & 0.989 & 0.912 & 0.985 & 0.931 & \textbf{0.996} & \textbf{0.971} & \textbf{1.000} & 0.998 \\
& & Prop. Best & 0.400 & 0.300 & 0.033 & 0.000 & \textbf{0.767} & \textbf{0.500} & \textbf{0.900} & \textbf{0.800} \\
\cline{2-11}
& \multirow{3}{*}{H-CLS (T=0.5)} 
  & Average & 0.960 & 0.804 & 0.952 & 0.879 & 0.933 & \textbf{0.937} & 0.898 & 0.979 \\
& & Max & 0.977 & 0.834 & 0.982 & 0.911 & 0.977 & 0.960 & 0.974 & 0.988 \\
& & Prop. Best & 0.000 & 0.000 & 0.000 & 0.000 & 0.100 & 0.467 & 0.000 & 0.133 \\
\cline{2-11}
& \multirow{3}{*}{H-CLS} 
  & Average & \textbf{0.981} & \textbf{0.900} & \textbf{0.982} & \textbf{0.928} & 0.970 & 0.935 & 0.992 & 0.984 \\
& & Max & \textbf{1.000} & \textbf{0.985} & \textbf{0.999} & \textbf{0.979} & \textbf{0.996} & \textbf{0.971} & \textbf{1.000} & \textbf{1.000} \\
& & Prop. Best & \textbf{0.600} & \textbf{0.700} & \textbf{0.967} & \textbf{1.000} & 0.133 & 0.033 & 0.100 & 0.067 \\
\hline
\multirow{9}{*}{150} 
& \multirow{3}{*}{E-CLS} 
  & Average & \textbf{0.980} & 0.883 & 0.954 & 0.873 & 0.964 & 0.934 & \textbf{0.989} & \textbf{0.982} \\
& & Max & 0.987 & 0.905 & 0.970 & 0.909 & 0.988 & 0.961 & \textbf{1.000} & \textbf{0.993} \\
& & Prop. Best & 0.400 & 0.300 & 0.100 & 0.000 & \textbf{0.667} & 0.333 & \textbf{0.833} & \textbf{0.733} \\
\cline{2-11}
& \multirow{3}{*}{H-CLS (T=0.5)} 
  & Average & 0.952 & 0.800 & 0.952 & 0.880 & 0.945 & 0.940 & 0.936 & 0.981 \\
& & Max & 0.979 & 0.820 & 0.979 & 0.900 & 0.983 & 0.959 & 0.984 & 0.987 \\
& & Prop. Best & 0.000 & 0.000 & 0.133 & 0.033 & 0.067 & \textbf{0.467} & 0.100 & 0.267 \\
\cline{2-11}
& \multirow{3}{*}{H-CLS} 
  & Average & 0.978 & \textbf{0.899} & \textbf{0.973} & \textbf{0.917} & \textbf{0.972} & \textbf{0.943} & 0.983 & 0.981 \\
& & Max & \textbf{0.998} & \textbf{0.966} & \textbf{0.994} & \textbf{0.958} & \textbf{0.994} & \textbf{0.972} & \textbf{1.000} & \textbf{0.993} \\
& & Prop. Best & \textbf{0.600} & \textbf{0.700} & \textbf{0.767} & \textbf{0.967} & 0.267 & 0.200 & 0.067 & 0.000 \\
\hline
\multirow{9}{*}{200} 
& \multirow{3}{*}{E-CLS} 
  & Average & 0.977 & 0.879 & 0.950 & 0.870 & 0.958 & 0.931 & \textbf{0.986} & \textbf{0.981} \\
& & Max & 0.985 & 0.902 & 0.972 & 0.915 & 0.991 & 0.966 & \textbf{1.000} & \textbf{0.993} \\
& & Prop. Best & 0.267 & 0.200 & 0.033 & 0.000 & 0.200 & 0.167 & \textbf{0.900} & \textbf{0.600} \\
\cline{2-11}
& \multirow{3}{*}{H-CLS (T=0.5)} 
  & Average & 0.952 & 0.802 & 0.960 & 0.885 & 0.947 & 0.940 & 0.935 & \textbf{0.981} \\
& & Max & 0.973 & 0.822 & 0.976 & 0.905 & 0.983 & 0.955 & 0.970 & 0.986 \\
& & Prop. Best & 0.000 & 0.000 & 0.100 & 0.033 & 0.067 & 0.067 & 0.067 & 0.400 \\
\cline{2-11}
& \multirow{3}{*}{H-CLS} 
  & Average & \textbf{0.982} & \textbf{0.909} & \textbf{0.975} & \textbf{0.920} & \textbf{0.980} & \textbf{0.956} & 0.973 & 0.980 \\
& & Max & \textbf{0.997} & \textbf{0.961} & \textbf{0.994} & \textbf{0.958} & \textbf{0.993} & \textbf{0.973} & \textbf{1.000} & \textbf{0.993} \\
& & Prop. Best & \textbf{0.733} & \textbf{0.800} & \textbf{0.867} & \textbf{0.967} & \textbf{0.733} & \textbf{0.767} & 0.033 & 0.000 \\
\hline
\end{tabular}%
}
\end{table}

We also evaluate the model performances under misspecified conditions. Specifically, we simulate 10 networks each of size $N=50$ and $N=100$ coming from different geometries and temperature settings: Euclidean (with temperature being 0.5), Hyperbolic with temperature being 0.01, and Hyperbolic with temperature being 0.5. We again compare the model performances using graph reconstruction accuracy and AUC. 

\begin{figure}[!ht]
    \centering
    \includegraphics[width=0.8\linewidth]{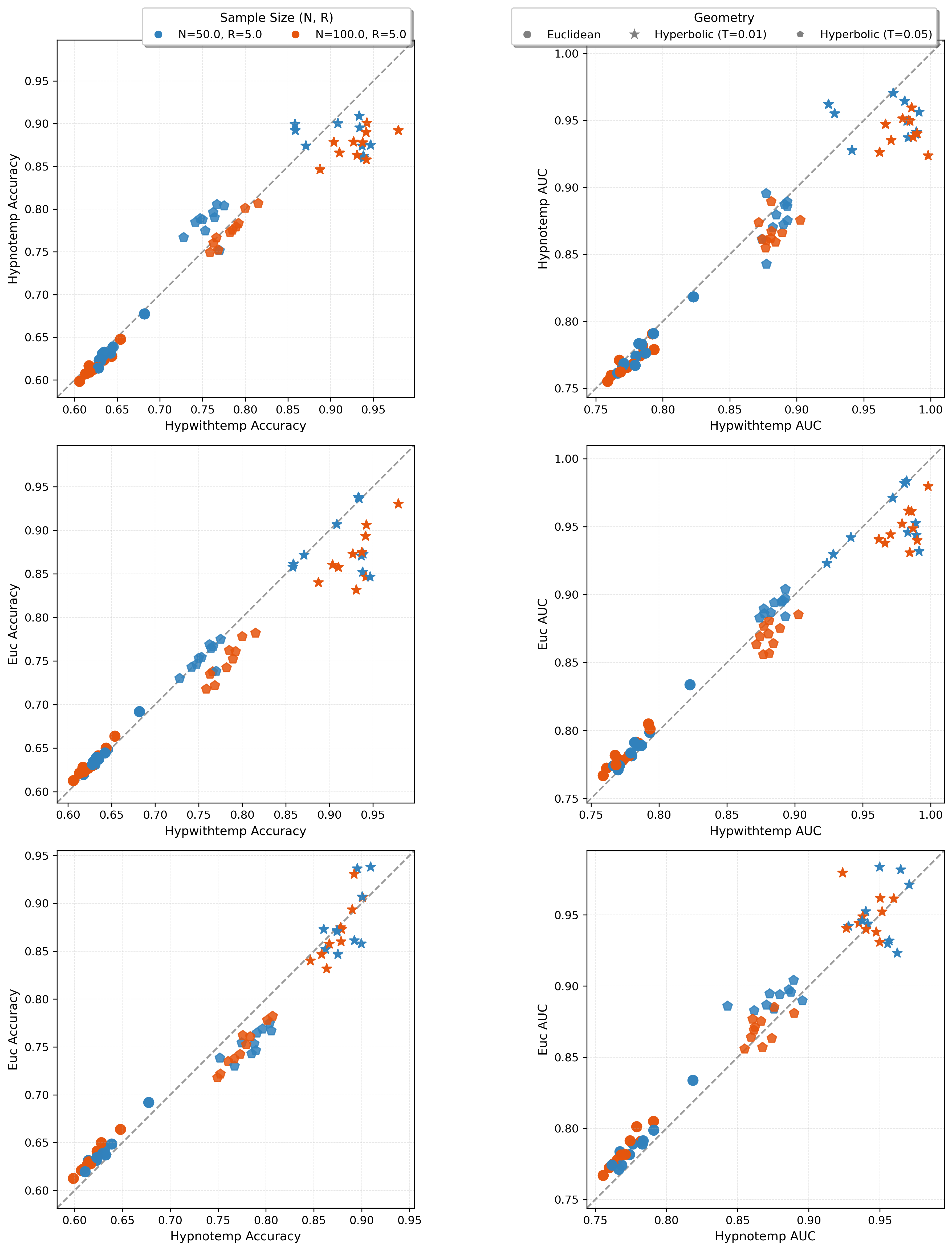}
    \caption{Pairwise comparison of model performance across networks simulated from different geometries and temperatures: Euclidean (circles), Hyperbolic geometry with $T=0.01$ (star), and Hyperbolic geometry with $T=0.5$ (pentagon). Each panel compares two models on accuracy (left column) and on AUC (right column) metrics. Points are colored by sample size $N$: blue for $N=50$ and orange for $N=100$. The dashed diagonal line represents equal performance between models; points above the line indicate the model on the y-axis performs better, while points below favor the model on the x-axis.}
    \label{fig:acrossgeom}
\end{figure}

As shown in Figure~\ref{fig:acrossgeom}, the hyperbolic model with learnable temperature (H-CLS) demonstrates robust performance across most settings, highlighting the flexibility gained from learning the temperature parameter. For networks simulated from hyperbolic latent spaces, regardless of whether the true temperature is low (0.01) or moderate (0.5), hyperbolic models with the estimable temperature generally achieve higher AUC and accuracy compared to the Euclidean model, with the performance gap widening as network size increases from $N=50$ to $N=100$. When comparing the two hyperbolic models, the learnable-temperature model (H-CLS) performs comparably or better across all data-generating scenarios, with its advantage more pronounced for data generated with low temperature.

For networks generated from the two-dimensional E-CLS model, where both hyperbolic models are substantially misspecified, the Euclidean model has some advantages, but the H-CLS model remains competitive. This indicates that temperature learning provides some degree of robustness even under severe model misspecification. Overall, when the underlying geometry is truly hyperbolic, both hyperbolic models consistently outperform the Euclidean model, with the performance advantage becoming more obvious as network size increases. 

In addition, we observe an important asymmetry in misspecification costs: using a Euclidean model on hyperbolic data incurs substantially larger performance penalties than using a hyperbolic model on Euclidean data. This suggests that hyperbolic models are more robust to geometric misspecification. This asymmetry may arise because hyperbolic space is locally Euclidean at small distances, allowing a hyperbolic model to accommodate near-Euclidean structure, but not vice versa. This finding further supports the practical utility of the H-CLS model, as it offers competitive performance even when the true geometry is Euclidean while providing substantial improvements when the underlying structure is indeed hyperbolic. 

\subsection{Variational inference model comparisons}

This subsection evaluates the scalability and empirical performance of our variational inference approach to model fitting. We fit the H-CLS and E-CLS models using the AEVB algorithm described in Section \ref{sec:vi}. The computational efficiency of this framework allows us to assess performance on networks of a much larger scale than is feasible with HMC.

We generate graphs from the H-CLS with true (low) temperature $T=0.01$.  We vary both network size with values of $N\in\{50, 100, 150, 200, 500, 1000, 5000\}$, and true hyperbolic radius $R\in\{3.0, 5.0, 7.0, 10.0\}$. We note that since \textit{a priori} $\alpha \sim N(R, 0.1)$, we are effectively considering the effect of true sparsity on model performance.  For each combination $N$ and $R$, we generate 10 network replicates using different random seeds to ensure the robustness of our findings. 

\begin{figure}[ht]
    \centering
    \includegraphics[width=0.9\linewidth]{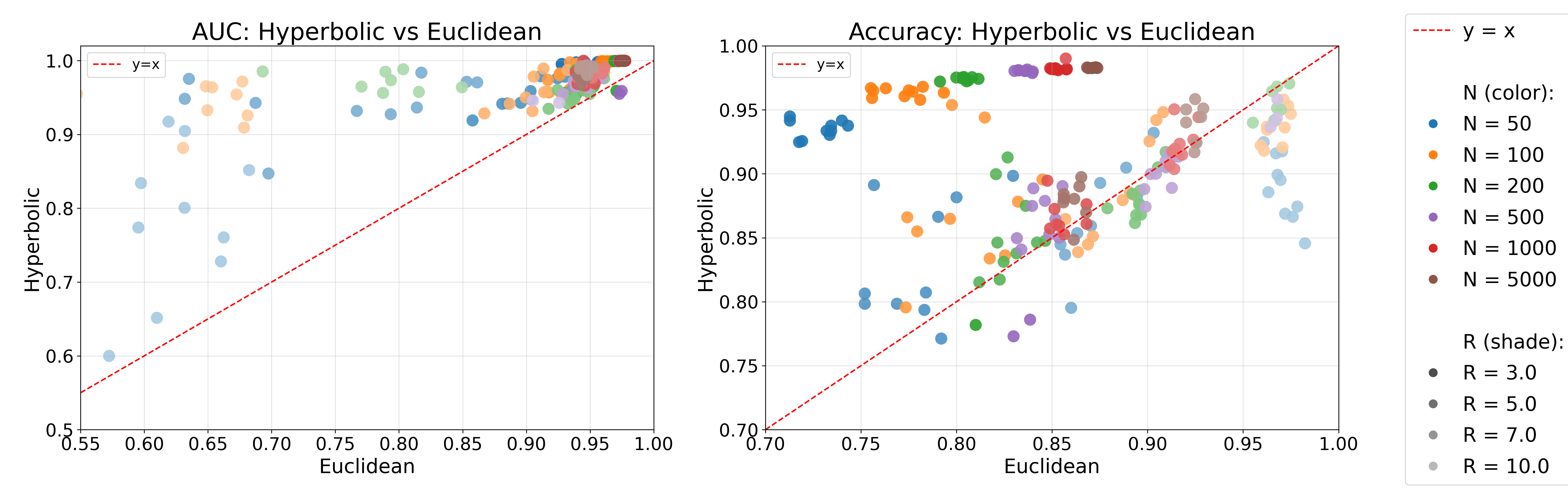}
    \caption{Comparison of H-CLS and E-CLS performance in a graph reconstruction task by accuracy and AUC. Each point represents the graph recovery performance metric as a function of model size ($N$) and true hyperbolic radius ($R$). Points above the dashed line $(x=y)$ indicate configurations where the H-CLS model outperformed the E-CLS model, and points below the line indicate the opposite.}
    \label{fig:acc_auc_euc_vs_hyp}
\end{figure}

Figure~\ref{fig:acc_auc_euc_vs_hyp} presents the primary results, showing the graph reconstruction AUC as a function of network size and hyperbolic radius $R$. Each point corresponds to a single simulation run for a specific combination of network size $N$ and true hyperbolic radius $R$. The performance of the misspecified E-CLS model is plotted on the x-axis against the performance of the H-CLS model on the y-axis. The diagonal $y=x$ line serves as a reference: points above this line indicate better performance by the correctly-specified H-CLS model.

The results demonstrate the clear advantage of the H-CLS model when it is the true generative model for the data. The H-CLS model consistently achieves a higher reconstruction accuracy and AUC than the E-CLS model across the large majority of the conditions. This advantage is particularly evident in the AUC panel. While both models perform well in some high-AUC areas (points clustered in the top-right corner), the H-CLS demonstrates a substantial performance gain in the more challenging scenarios where the E-CLS model struggles (points in the lower and middle range of the x-axis). This suggests that as the network structure becomes more complex (greater radius/sparsity), the hyperbolic geometry provides a crucial advantage for accurately capturing the network's true tree-like topology. Similarly, the same trend is visible in the accuracy panel. While the H-CLS model consistently outperforms the E-CLS alternative, both models achieve high accuracy scores across most simulation settings. This is expected in sparse graphs where correctly identifying the numerous non-edges can lead to high accuracy. The AUC, being a rank-based metric, therefore serves as a more discriminative measure of a model's ability to accurately predict edges than the absence of edges.  

\begin{figure}[ht]
    \centering
    \includegraphics[width=0.9\linewidth]{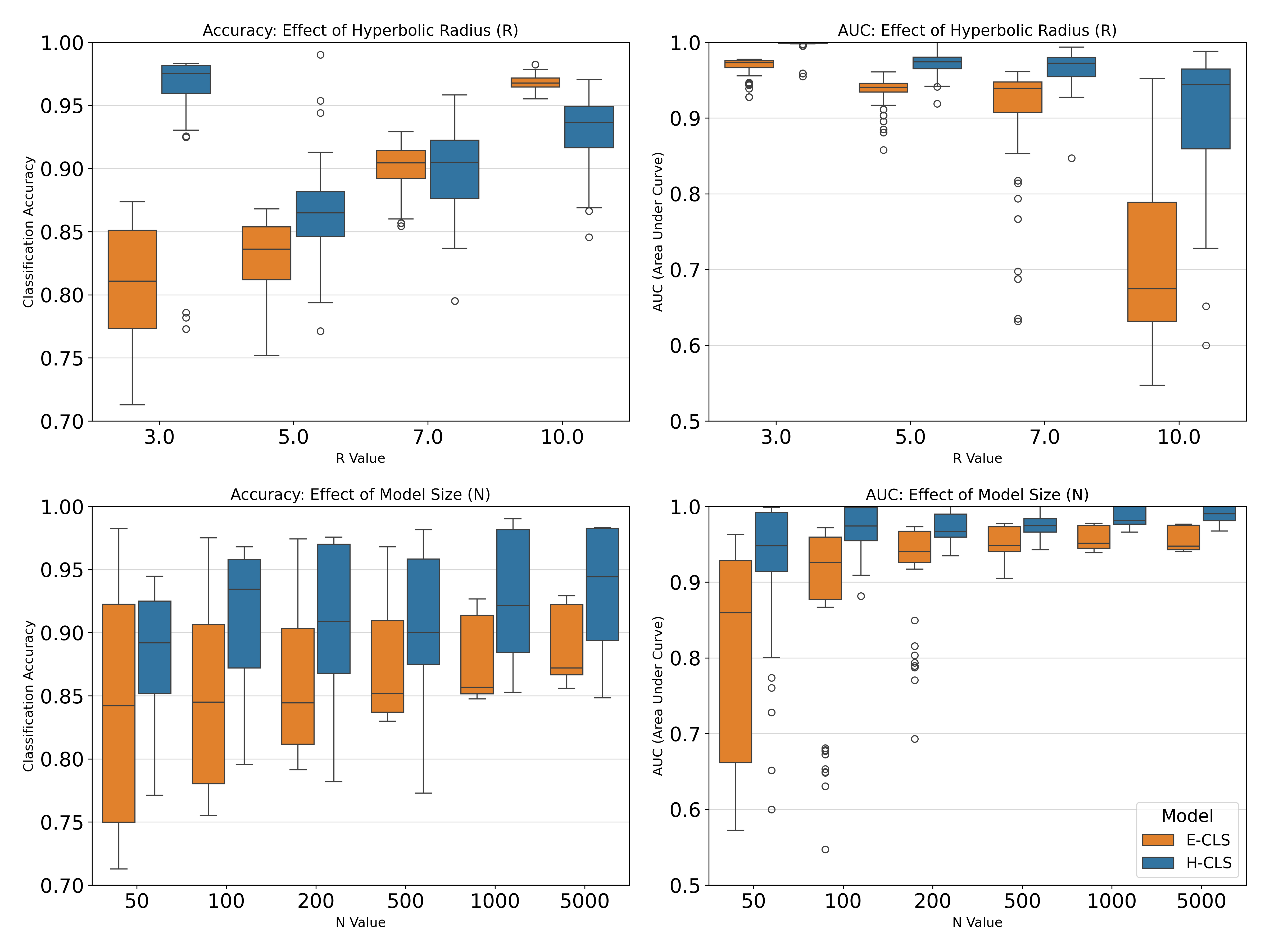}
    \caption{Comparison of the performance of H-CLS and E-CLS on a network reconstruction task across different radius $R$ and network size $N$. Box plots illustrate the marginal effect of the two key parameters on graph recovery across 40 different configurations.}
    \label{fig:boxplot_hyp_vs_euc}
\end{figure}

To further investigate the performance gap between the correctly and incorrectly specified models, Figure~\ref{fig:boxplot_hyp_vs_euc} breaks down the results from our simulation study, examining the effects of the hyperbolic radius, $R$, and the network size, $N$, independently. The bottom panels illustrate the effect of network size. As $N$ increases, the performance of both models consistently improves and their variance decreases, with H-CLS AUC approaching 1.0 for networks of 500 nodes or more. This demonstrates that the correctly-specified model's performance improves--leveraging the latent hyperbolic geometry--as number of nodes increases. While the incorrectly-specified E-CLS model's performance also trends upward with $N$, it does so at a much slower rate and with higher variance, never achieving the near-perfect graph reconstruction observed with the H-CLS model.  

\subsection{Real data example}
\label{sec:examples}

To demonstrate the practical utility of our H-CLS model beyond simulated data, we evaluate its performance on four network data sets representative of a diverse collection of application domains (social, biological, and energy infrastructure). As shown in Table~\ref{tab:network}, these networks exhibit varying structural properties that test different aspects of our modeling framework: the Facebook--Simmons college network for students graduating between 2006-2009 (N=1168))~\citep{facebook} represents dense social connectivity with high clustering (0.32); the Caenorhabditis elegans (C. elegans) neural network (N=297)~\citep{watts1998collective} captures biological organization with moderate clustering (0.292); the Dolphin social network (N=62)~\citep{watts1998collective} provides a smaller-scale test case with natural community structure; and the Western States Power Grid of the United States network (N=4941) represents sparse infrastructure topology with notably low clustering (0.038). We fitted both the H-CLS and E-CLS to the four data sets using the AEVB algorithm. We employed identical architectures, two-layer graph convolution network encoders with 128-dimensional hidden representations, to ensure fair comparison, with one exception for the Power Grid network. For this network, the H-CLS model required increased capacity (1024 hidden dimensions) to provide a result better than random guessing, likely due to the extreme sparsity. For each network, we evaluate graph reconstruction performance using both the AUC and accuracy metrics, as described in Section~\ref{sec:infer}.

\begin{table}[ht]
    \centering
    \small
    \begin{tabular}{l *{7}{c}} 
        \toprule
        \textbf{Dataset}  & \textbf{c(G)} & \textbf{\# Triangles} & \textbf{C(G)} & \textbf{\# of Nodes} & \textbf{Edge Density} \\
        \midrule
        Facebook--Simmons  & 23292 & 121514 & 0.318 & 1168 & 0.036 \\
        C. Elegans  & 1852 & 3241 & 0.292 & 62 & 0.084 \\
        Dolphin  & 98 & 98 & 0.259 & 297 & 0.049 \\
        Power Grid  & 1654 & 651 & 0.038 & 4941 & 0.0005 \\
        \bottomrule
    \end{tabular}
    \caption{Network summary statistics for the four examples in Section \ref{sec:examples}.  $c(G)$ and $C(G)$ are the circuit rank and global clustering coefficients, respectively, as defined in Section \ref{sec:gap}.} 
    \label{tab:network}
\end{table}

\begin{table}[ht]
    \centering
    \begin{tabular}{l *{7}{c}} 
        \toprule
        \textbf{Dataset} & \textbf{H-CLS AUC} & \textbf{H-CLS Accuracy} &\textbf{E-CLS AUC} &\textbf{E-CLS Accuracy} \\
        \midrule
        Facebook--Simmons & 0.895 & 0.944 & 0.886 & 0.937 \\
        C. Elegans & 0.897 & 0.931 & 0.853 & 0.914 \\
        Dolphin & 0.867 & 0.902 & 0.812 & 0.867 \\
        Power Grid & 0.769 & 0.999 & 0.754 & 0.999 \\
        \bottomrule
    \end{tabular}
    \caption{Summary of the performance of H-CLS and E-CLS models on real data examples introduced in Section \ref{sec:examples}.}
    \label{tab:recon}
\end{table}

Table~\ref{tab:recon} presents graph reconstruction performance across the four real-world networks, comparing H-CLS and E-CLS models. The H-CLS model consistently outperforms its Euclidean counterpart in terms of AUC, with improvements ranging from 1.5\% (Power Grid) to 5.5\% (Dolphine network). The performance gains are most pronounced for networks with stronger clustering structure: the Dolphin social network and C. elegans neural network show 5.5\% and 4.4\% AUC improvements respectively, while the Power Grid, which has more uniform degree distribution and less hierarchical structure, shows modest gains. Classification accuracy follows similar patterns, with hyperbolic models achieving 2-4\% improvements for biological and social networks. These results confirm that hyperbolic geometry provides quantitative advantages for graph reconstruction, particularly in networks with inherent hierarchical organization, while the consistent improvements across diverse domains suggest broad applicability of approach. 

\subsubsection{Example: uncovering structure in the Facebook 100 network}

To further demonstrate the advantage of our model, we present analysis of the an Facebook--Simmons College social network data set ~\citep{facebook}. The data set consists of $N=1168$ students (nodes) who graduated from Simmons College between 2006 and 2009 and their undirected Facebook friendship relationships (edges). We fitted both the E-CLS and H-CLS models to these data using the AEVB algorithm.  

\begin{figure}[htbp]
    \centering
    
    \subfloat[H-CLS embedding colored by graduation year.]{\label{fig:hyperbolic-year}\includegraphics[width=0.45\textwidth]{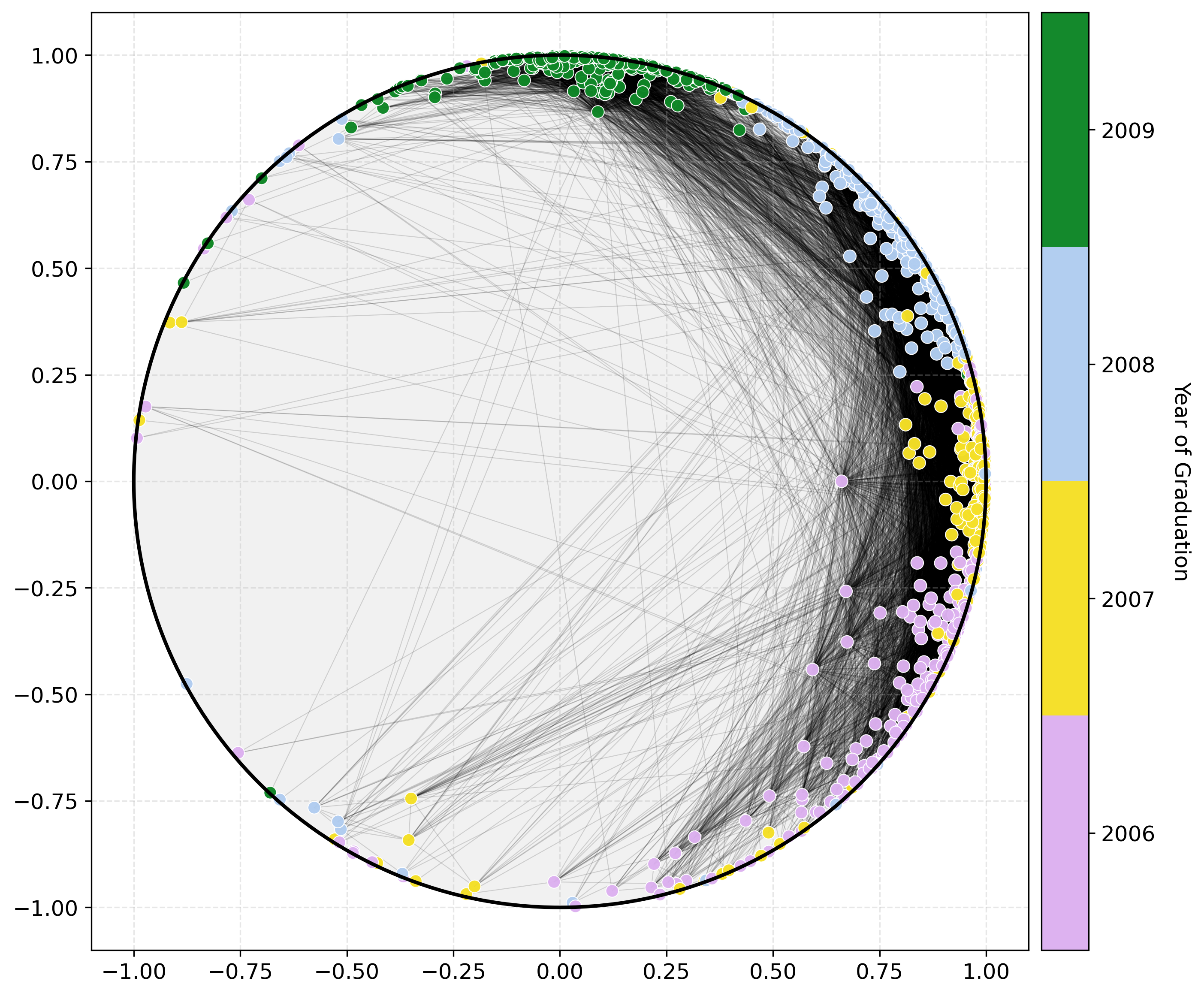}}
    \hfill 
    \subfloat[E-CLS embedding colored by graduation year.]{\label{fig:euclidean-year}\includegraphics[width=0.41\textwidth]{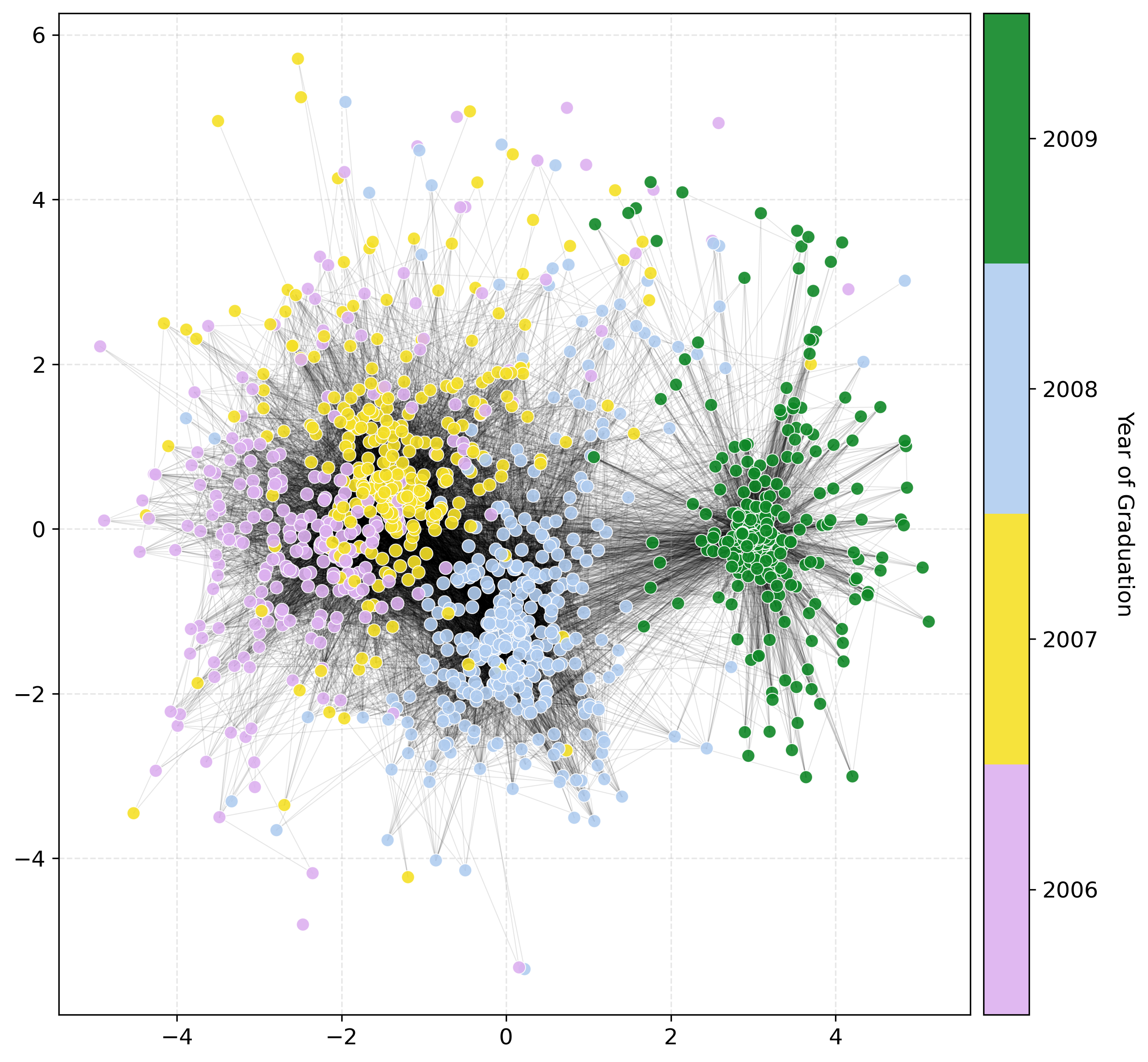}}

    \caption{Comparison of H-CLS and E-CLS embeddings for the Facebook--Simmons dataset colored by graduation year. The left column shows the H-CLS model's estimated embeddings, and the right column shows the E-CLS model's estimated embeddings.}
    \label{fig:facebook}
\end{figure}

Figure~\ref{fig:facebook} displays the posterior mean embeddings from fitting both models using the AEVB algorithm to the Facebook--Simmons college social network from class 2006 to 2009 across N=1168 students~\citep{facebook}.  

For visualization and interpretation, we construct estimated embedding for both models by by computing point estimates of the latent positions. For the H-CLS model, we took the estimated the variational parameters, $\left\{\mu_{ir}\right\}_{i=1}^{N}$ and $\left\{\mu_{i\theta}\right\}_{i=1}^{N}$ and transformed them through the nonlinear mapping described in Section~\ref{sec:vi} to obtain hyperbolic coordinates for each node. To ensure stable and visually comparable visualizations, we then applied a canonical Procrustes rotation to the embeddings analogously to \citealp{hoff2002latent}'s proposal. For the H-CLS-generated embeddings, we rotated the Poincare disk to place the most central node (minimum radius) at angle zero, then reflected across the horizontal axis if needed to concentrate the weighted angular mass in the upper half-plane. For the Euclidean case, we took the estimates of the latent position  parameters and directly applied the Procrustes rotation to them.  Both sets of rotated embeddings are displayed in Figure~\ref{fig:facebook}. In both cases, color is used to distinguish the students' graduate year, a factor not used in model fitting but rather one that seems relevant to the friendship experiences of students in the school.  
 
While Table~\ref{tab:recon} shows the H-CLS model achieves modest quantitative improvements in the graph reconstruction task, the embeddings reveal a more substantial qualitative advantage: the H-CLS model recovers interpretable latent structure that directly corresponds to the social organization of the network (i.e., students sorted by graduate year). This additional interpretability, provides scientific insights beyond predictive performance alone.  

The model discovers a clear temporal organization, with graduation cohorts arranged sequentially in angular space around the Poincaré disk. Students from the same graduation year form distinct, cohesive clusters, while the angular ordering reflects the temporal sequence of cohorts. This organization emerges purely from network topology, as graduation year was not provided to the model during inference. The angular coordinate thus serve as a latent temporal axis, with the 2006-2009 cohorts appearing in order.   

The edge density between adjacent cohorts reflects their temporal overlap on campus. Notably, the 2008 and 2009 cohorts exhibit particularly dense inter-group connections, consistent with their three-year co-presence at the university. Conversely, minimal direct connections exist between the 2006 and 2009 cohorts, who had little to no temporal overlap. This gradient of cross-cohort connectivity is naturally captured by the hyperbolic distance metric, where angular separation encodes social distance. A small subset of 2007 nodes appears displaced toward the 2006 cluster, perhaps representing students with delayed graduation who maintained stronger ties to their original cohort.  



In contrast, the E-CLS estimated embeddings fail to recover this temporal organization. Panel (b) reveals substantial mixing of graduation cohorts, with only the 2009 class achieving partial spatial separation but the 2006-2008 cohorts form an intermingled mass in the left region. 

The preferable performance of H-CLS model stems from how hyperbolic geometry separates roles between coordinates. The angular coordinate encodes temporal progression and group membership, arranging cohorts sequentially around the disk's circumference, while the radial dimension captures hierarchical depth and centrality within the network. These complimentary roles enable the model to simultaneously preserve local clustering and global temporal organization.  

\section{Discussion}\label{sec:discussion}

In this work, we have revisited the hyperbolic continuous latent space model, focusing on the central and often overlooked role of the temperature parameter $T$. We have argued that the common practice of fixing or omitting the temperature parameter obscures a core modeling degree of freedom, the extent to which a network's connectivity is governed by its latent geometry versus stochastic noise. Our primary contribution is formalizing the estimation of $T$ within a Bayesian framework, yielding both improved graph reconstruction and, more importantly, geometrically interpretable embeddings that reveal latent hierarchical structure. 

This improvement highlights a key insight: the degree to which a network adheres to a pure geometric structure is itself a property to be learned from the data. Real-world networks are not perfect geometric graphs: they contain noise, random connections, and deviations from pure hierarchy. The temperature parameter directly models the degree of this deviation, an idea precedent in Euclidean~\cite{rastelli2016properties} and hyperspherical~\cite{mccormick2015latent} latent space models, but not previously studied in hyperbolic settings. By fixing $T$, previous models implicitly assumed a fixed level of randomness, whereas our approach allows the model to adapt to the specific structural properties of the observed graph, from highly tree-like ($T\longrightarrow 0$) to more densely clustered. 

To make this modeling framework more computationally feasible, we developed both an HMC algorithm for full posterior inference and an AEVB algorithm that scales to networks with thousands of nodes, whose prior structure further highlights the importance of a flexible generative process in achieving high performance. 

Our empirical validation across both simulated and real-world networks confirms the practical value of this approach. Notably, the magnitude of improvement correlates with network clustering coefficient, suggesting hyperbolic geometry provides greatest benefit for networks with strong community structure. Beyond quantitative improvements, our analysis reveals a distinct qualitative advantage: geometric interpretability. The Facebook-Simmons embeddings demonstrate that hyperbolic geometry naturally recovers meaningful latent structure: graduation years form distinct angular sectors while maintaining hierarchical radial organization, despite having no access to node attributes during training. It echoes findings by~\cite{hoff2002latent}, where latent positions on a sphere separated students by gender without explicit covariate information. This emergent structure aligns with expected social dynamics where temporal cohorts and popularity hierarchies shape network formation. In contrast, the Euclidean embeddings produce less interpretable configurations. 

Several future research directions remain. First, while we have established the practical importance of inferring $T$, the partial identifiability between $T$, the sparsity parameter $\alpha$ and the latent scale, $R$, remains a theoretical challenge. Relatedly, our choice to relax the constraint $\alpha=R$ used in~\cite{krioukov2010hyperbolic} to $\alpha \sim 
\text{Normal}(R, \sigma)$ introduces additional flexibility; the extent to which this relaxation affects inference and predictive performance deserves further study. Further work could explore more sophisticated prior structures or reparameterizations to mitigate these issues. Second, our work has focused on general static networks. Extending this framework to dynamic networks, where node positions and perhaps even the global temperature evolve over time, is a critical next step for modeling real-world systems.

\newpage
\bibliography{references}
\bibliographystyle{abbrvnat}

\end{document}